\documentclass[12pt]{article}

\usepackage{amssymb,amsmath}
\usepackage{graphicx}

\newcommand{\be}{\begin{equation}}
\newcommand{\ee}{\end{equation}}
\newcommand{\Dlt}{\Delta}
\newcommand{\dlt}{\delta}
\newcommand{\prt}{\partial}
\newcommand{\br}{{\bf r}}
\newcommand{\bk}{{\bf k}}

\newcommand{\ba}{{\bf a}}
\newcommand{\bu}{{\bf u}}
\newcommand{\bp}{{\bf p}}
\newcommand{\bbe}{{\bf e}}
\newcommand{\bS}{{\bf S}}
\newcommand{\bt}{\beta}
\newcommand{\vp}{\varphi}
\newcommand{\ep}{\varepsilon}
\newcommand{\al}{\alpha}
\newcommand{\ra}{\rightarrow}
\newcommand{\sgm}{\sigma}

\newcommand{\gm}{\gamma}
\newcommand{\om}{\omega}
\newcommand{\Om}{\Omega}
\newcommand{\Gm}{\Gamma}
\newcommand{\dgr}{\dagger}
\newcommand{\lbd}{\lambda}
\newcommand{\Lbd}{\Lambda}

\newcommand{\rgl}{\rangle}
\newcommand{\lgl}{\langle}

\begin{document}

\begin{center}
{\Large{\bf
Difference in Bose-Einstein condensation of conserved and 
unconserved particles} \\ [5mm]
V.I. Yukalov} \\ [3mm]

{\it Bogolubov Laboratory of Theoretical Physics, \\
          Joint Institute for Nuclear Research, Dubna 141980, Russia}
\end{center}

\vskip 3cm

{\bf Keywords}: Bose-Einstein condensation, trapped atoms, optical lattices, 
conserved and unconserved particles, quasiparticles, vibrating optical lattices, 
double-well lattices, condensation of collective excitations, phonons, magnons, 
Schwinger bosons, slave bosons, dimer lattices, condensation of singletons and 
triplons, strongly nonequilibrium condensates

\vskip 3cm

\begin{abstract}  

The peculiarities in the Bose-Einstein condensation of particles and 
quasiparticles are discussed. The difference between the condensation of 
conserved and unconserved particles is analyzed. A classification of 
quasiparticles is given. The emphasis is made on the ability of particles 
and quasiparticles to condense. Illustrations include: general Bose-condensed 
atomic systems, such as ensembles of trapped atoms, Bose gases with conserved 
and unconserved number of atoms, vibrating atoms in double-well lattices, 
Holstein-Primakoff magnons, Schwinger bosons, slave bosons, and the 
condensation of singletons and triplons. The basic difference is that the 
system of particles, whose total number is conserved, can form equilibrium as 
well as nonequilibrium condensates, while unconserved particles can condense 
only in a nonequilibrium system subject to external pumping supporting the 
density of these particles sufficient for their condensation. The examples 
of such a nonequilibrium condensation of unconserved particles are the 
Bose-Einstein condensation of excitons, polaritons, and photons. Elementary
collective excitations, such as bogolons and phonons, being self-consistently 
defined, do not condense. Magnons cannot condense in equilibrium systems. 
Controversies, existing in literature with regard to the Bose-Einstein 
condensation of some quasiparticles, are explained. Pushing a system out of 
equilibrium may favor the condensation of unconserved quasiparticles, but 
suppresses the condensate fraction of conserved particles.

\end{abstract} 

\newpage

\section{Introduction}
 
Bose-Einstein condensation has been a topic of intensive studies, both 
experimental and theoretical. There exists numerous literature on the problem, 
which can be found in the books [1-4] and review articles [5-18]. A variety of 
experiments have been performed with cold trapped atoms. In addition to atoms, 
whose total average number can be well controlled, hence, is conserved, one often
considers Bose-Einstein condensation of unconserved particles, e.g., corresponding 
to excited bound states or collective excitations. The known examples are the
Bose-Einstein condensation of excitons [19-23], polaritons [24-27], and, recently, 
photons [28].

Sometimes controversy arises, accompanying the discussion on the feasibility of 
realizing the Bose-Einstein condensation of unconserved particles. A very 
illustrative case of such a controversy is the discussion on the possibility 
of magnon condensation. In several works (e.g. [29-32]), the authors claimed the 
existence of magnon condensation in some equilibrium magnetic materials. Because 
of the known mapping [33,34] of spin systems to Bose gas, the existence of magnon 
condensation would mean the occurrence of Bose condensation for the collective 
excitations (bogolons) of Bose systems. Is it feasible that bogolons would 
condense, may be in a nonequilibrium Bose gas [35] in an external alternating 
potential? The problems of whether bogolons and magnons could condense are
closely related. Mills [36] stressed that magnon condensation in equilibrium 
systems is impossible. Why this is impossible and could magnons condense in 
nonequilibrium systems? More generally, could elementary collective excitations
become condensed and, if so, when?     

Understanding the necessary conditions, when particles and quasiparticles could 
form Bose-Einstein condensates, is a general problem that has become of great 
importance due to high current interest to Bose-condensed systems [1-18]. In the 
present paper, a careful analysis is given of such necessary conditions of 
condensation. The consideration is illustrated by the cases of the most common 
types of quasiparticles whose classification is also given. The suggested analysis 
resolves some controversies arising with regard to the possibility of Bose-Einstein 
condensation of quasiparticles.

\section{Particles versus quasiparticles}

Traditionally, one distinguishes elementary particles and quasiparticles. The 
distinction, however, can differ depending on the considered physical applications.

In particle physics, an elementary particle, or fundamental particle, is a particle 
not known to have substructure; that is, it is not known to be made up of smaller 
particles. If an elementary particle truly has no substructure, then it is one of 
the basic building blocks of the universe from which all other particles are made. 
In the Standard Model, the quarks, leptons, and gauge bosons are elementary 
particles [37].

In statistical physics, one calls an elementary particle that whose structure is
either absent or of no importance for the considered application. So, hadrons (mesons 
and baryons, such as the proton and neutron) and even whole atoms and molecules may 
be regarded as elementary particles, provided that their structure, if it exists,
can be neglected for the studied phenomena.   

The notion of quasiparticles is not well defined. One often calls quasiparticles 
the dressed particles whose properties are changed by the surrounding medium. 
For example, as an electron travels through a semiconductor, its motion is disturbed 
in a complex way by its interactions with all of the other electrons and nuclei. 
However it approximately behaves like an electron with a different mass traveling 
unperturbed through free space. This "electron" with a different mass is called an 
electron quasiparticle or Landau quasiparticle. Various single-particle and 
collective excitations are also termed quasiparticles. To make the consideration
better defined, we give below a classification of different types of quasiparticles.

\subsection{Dressed particles}

Dressed particles are those that can be treated as elementary, but with the 
properties that are changed by surrounding, making them different from their 
properties in vacuum. Examples are: electron in a solid, electron or ion in 
plasma, proton or neutron inside a nucleus, polaron (electron with polarization of 
surrounding ions of a crystalline lattice), polariton (photon in a medium). 
Sometimes composite particles can mathematically be treated as elementary, such as
atoms and molecules. A stable bipolaron (bound pair of two polarons) can be 
considered as one Bose particle. Strictly speaking, in quantum field theory, all 
particles interact with vacuum, being in that sense all dressed. The dressed 
particles can be bosons or fermions depending on their spin.

\subsection{Particle excitations}

Some quasiparticles behave as dressed particles, but differ from them by the 
necessity of supplying energy to the system for their creation. In that sense,
such particle excitations are not stable and decay in an equilibrium system.
For instance, a hole in the valence band of a semiconductor is a lack of an 
electron that has been excited to the conductance band. A hole is a fermion 
quasiparticle. Another example is an exciton in a semiconductor, that is a 
bound state of an electron and a hole, requiring for its creation photon 
excitation. Excitons can be treated as bosons. The number of excitations is 
not conserved, depending on the intensity of the applied external pumping.

\subsection{Collective excitations}  

Collective excitations describe quantized fluctuations of an order parameter.
Examples are: phonon (quant of mass density fluctuations), plasmon (quant of
charge density fluctuations), magnon (quant of spin fluctuations), bogolon 
(quant of density fluctuations in a Bose-condensed system). Such collective 
excitations are usually described by linear equations characterizing small 
deviations from a stationary state, because of which they are termed 
elementary collective excitations. All these collective excitations are bosons. 
They are not conserved.

\subsection{Nonlinear waves}

Another type of excitations can also be formed by the system as a whole, being 
in that sense also collective, but, contrary to the elementary collective 
excitations, being described by nonlinear equations corresponding to large 
deviations from a stationary state. In classical and quasiclassical picture
such excitations are named nonlinear waves. As examples, we can mention 
solitons [38,39], bions (two bound solitons) [40,41], triple solitons (three 
bound solitons) [42], quantum vortices [17], and topological coherent modes 
[43-54]. Being quantized, such nonlinear excitations can be interpreted as 
quasiparticles. Usually they are not stable, except some one-dimensional 
solitons, arise only in nonequilibrium systems, and are not conserved.

\subsection{Auxiliary quasiparticles}

In quantum theory, particles and quasiparticles are represented by field 
operators of creation and annihilation. The operators can be transformed 
resulting in the appearance of new operators that can correspond to other
physical quasiparticles. Thus, the field operator of uncondensed particles 
in a Bose-condensed system can be decomposed into the linear combination of 
operators corresponding to collective excitations, bogolons. But it is not 
necessary that any operator transformation would result in new operators 
corresponding to some physical particles or quasiparticles. A transformation
can be just a mathematical formula, convenient for calculations, but not 
necessarily related to new physical objects. Nevertheless, it is customary
to associate each field operator, occurring in operator transformations, 
with a quasiparticle, though this can be merely a fictitious particle being
related to no real physical object. Such auxiliary fields and quasiparticles
occur, e.g., in the Hubbard-Stratonovich transformation [55]. Other 
examples will be given below. Auxiliary particles can be either bosons or 
fermions, or even anyons, having intermediate fractional statistics [56].

\section{Condensation of cold atoms}

There are two ways of treating Bose-Einstein condensation of particles whose
total number is fixed. It is possible to consider a finite system of $N$
particles in volume $V$. In a finite system, strictly defined phase transitions
are absent. Following a kind of the Penrose-Onsager scheme [57], one needs 
to find the largest eigenvalue of the single-particle density matrix, which 
represents the number of condensed particles $N_0$. If the value 
$n_0 \equiv N_0/N$ in thermodynamic limit remains finite, one says that 
Bose-Einstein condensation takes place. In that case, the order index of the
single-particle density matrix becomes unity [58-60]. This way is extremely 
complicated and, for realistic models, can be realized only by means of 
powerful computers.  

Another way takes into account that, though strictly defined phase transitions, 
including Bose-Einstein condensation, occur only in thermodynamic limit, but, 
if the number of particles in the considered system is sufficiently large, such
that $N \gg 1$, then one can use the relations typical of infinite systems,
where phase transitions are well defined. This essentially simplifies the 
consideration. In thermodynamic limit, Bose condensation is accompanied by 
spontaneous breaking of global gauge symmetry. Moreover, the symmetry breaking 
is necessary and sufficient for Bose-Einstein condensation [11]. In a finite
system, the symmetry breaking is {\it asymptotic}, in the sense that it is 
approximate for finite $N$, and becomes exact for $N \ra \infty$. For a large 
number of particles $N \gg 1$, the symmetry can be considered as broken. Then 
it is reasonable to take into account the symmetry breaking explicitly, which 
makes the theory more transparent and calculations treatable. 

In the case of Bose condensation, the most convenient method of symmetry 
breaking is by means of the Bogolubov shift [61,62], when the Bose field 
operator is represented as the sum
\be
\label{1}
 \hat\psi (\br,t) = \eta(\br,t) + \psi_1(\br,t) \;  ,
\ee
in which the first term is the condensate wave function and the second term 
is the field operator of uncondensed atoms. The latter two quantities are 
orthogonal to each other,
\be
\label{2}
 \int \eta^*(\br,t) \psi_1(\br,t)\; d\br = 0  \; .
\ee
Uncondensed atoms are considered as normal, such that the statistical average 
of the operator of uncondensed atoms is zero:
\be
\label{3}
 \lgl \psi_1(\br,t) \rgl = 0 \;  .
\ee
Then the average of the total field operator (1),
\be
\label{4}
\lgl \hat \psi(\br,t) \rgl = \eta(\br,t) \;   ,
\ee
gives the condensate wave function that can be treated as the order parameter.
Its modulus squared yields the condensate density
\be
\label{5}
 \rho_0(\br,t) = | \; \eta(\br,t) \; |^2 \;  ,
\ee
which is normalized to the number of condensed atoms
\be
\label{6}
 \int | \; \eta(\br,t) \; |^2\; d\br = N_0(t) \;  .
\ee
Another normalization condition gives the number of uncondensed atoms
\be
\label{7}
 \lgl \hat N_1(t) \rgl = N_1(t)  
\ee
that is the average of the number operator
\be
\label{8}
  \hat N_1(t) \equiv \int \psi_1^\dgr(\br,t) 
\psi_1(\br,t) \; d\br .
\ee
The sum of the numbers of condensed, (6), and uncondensed, (7), atoms gives
the total number of atoms
\be
\label{9}
 N_0(t) + N_1(t) = N \;  .
\ee
Generally, the total number of atoms could depend on time, when there is the 
input of atoms or their loss. But in any case, this number is fixed at each 
moment of time by the given external conditions and cannot be created inside 
the system. 

Equation (3) can be rewritten in the standard form of a statistical condition
by introducing the operator
\be
\label{10}
  \hat\Lbd(t) \equiv \int \left [ \lbd(\br,t) \psi_1^\dgr(\br,t) +
\lbd^*(\br,t) \psi_1(\br,t) \right ] \; d\br \; .
\ee
Then Eq. (3) is equivalent to the condition 
\be
\label{11}
 \lgl \hat\Lbd(t) \rgl = 0 \;  .
\ee

The above formulas are defined for an arbitrary system, equilibrium or 
nonequilibrium, provided that the condensate does exist, that is, when the 
{\it condition of condensate existence}
\be
\label{12}
\lim_{N\ra\infty} \; \frac{N_0(t)}{N} \; > \; 0
\ee
is valid. The system as such can be finite, but the thermodynamic limit 
in Eq. (12) serves as the check for the accurate definition of the condensate
existence. This condition imposes constraints on the behavior of the  
distribution of uncondensed atoms and the spectrum of collective excitations 
[16,18] for an equilibrium system. The distribution of atoms $n_k$ over the 
quantum multi-indices $k$ is to be such that, if the condensation is into 
the state labelled by $k_0$, then
\be
\label{13}
  n_k \ra \infty \qquad ( k \ra k_0 ) \; .
\ee
This is equivalent to the requirement that the spectrum $\varepsilon_k$ be 
gapless:
\be
\label{14}
 \ep_k \ra 0 \qquad ( k \ra k_0 ) \;  .
\ee

Another important point is that the cases, when the condensate exists or does 
not exist require different mathematics [63]. When the condensate exists, hence 
condition (12) holds true, then the action of all operators is defined on the
Fock space ${\cal F}(\psi_1)$ generated by the field operator $\psi_1$. But
when there is no condensate, so that the condensate function $\eta(\br,t)$ and 
the condensate density (5) are zero, then the theory is defined on the Fock 
space ${\cal F}(\psi)$ generated by the field operator $\psi$, without any shift.
These two spaces are orthogonal to each other [63]. It is incorrect to work in 
the space ${\cal F}(\psi)$ and then make the Bogolubov shift (1) passing to the 
space ${\cal F}(\psi_1)$. This is the standard mistake. As soon as there happens
Bose-Einstein condensation, one has to work in the space ${\cal F}(\psi_1)$.

Let $\hat{H}[\eta,\psi_1]$ be the energy Hamiltonian for the system. Because 
of the existence of three statistical conditions (6), (7), and (11), the grand 
Hamiltonian must contain three additional terms:
\be
\label{15}
 H = \hat H [ \eta, \; \psi_1 ] - \mu_0 N_0(t) -
\mu_1 \hat N_1(t) - \hat\Lbd(t) \;  .
\ee
Here $\mu_0$ and $\mu_1$ are the Lagrange multipliers guaranteeing the 
normalizations (6) and (7). And $\lbd(\br,t)$ is the Lagrange multiplier 
cancelling the linear in $\psi_1({\bf r}, t)$ terms in order to satisfy 
condition (11). The grand Hamiltonian (15) allows one to correctly define a 
representative statistical ensemble uniquely describing the Bose-condensed 
system [63-67]. 

The equations of motion can be represented by the variational derivatives 
of the grand Hamiltonian over $\eta$ and $\psi_1$, which is equivalent to the 
Heisenberg equations of motion [18,68]. Thus, the equation for the condensate 
wave function reads as 
\be
\label{16}
 i \; \frac{\prt}{\prt t} \; \eta(\br,t) = \left \lgl
\frac{\dlt H}{\dlt\eta^*(\br,t) } \right \rgl \;  .
\ee
In the case of an equilibrium system, this reduces to the equation
\be
\label{17}
 \left \lgl \frac{\dlt H}{\dlt\eta^*(\br) } \right \rgl  = 0 \; .
\ee
And, if the equilibrium system is uniform, the previous equation becomes 
\be
\label{18}
 \left \lgl \frac{\prt H}{\prt N_0 } \right \rgl = 0 \;  .
\ee 

Evolution equations, together with the related initial conditions, define
the functions $N_0(t,\mu_0,\mu_1)$ and $N_1(t,\mu_0,\mu_1)$. Because of 
normalization (9), the multipliers $\mu_0$ and $\mu_1$ are connected with 
each other and, generally, can depend on time. The condition of the gapless 
spectrum, for a nonequilibrium system, reads as
\be
\label{19}
 \ep_k(t) \ra 0 \qquad (k \ra k_0 ) \;  .
\ee
These equations define the multipliers $\mu_0(t,N)$ and $\mu_1(t,N)$.

The time-dependent solutions should tend to the corresponding stable stationary 
solutions $N_0 = N_0(\mu_0,\mu_1)$ and $N_1 = N_1(\mu_0,\mu_1)$ that, according 
to sum (9), satisfy the normalization $N_0 + N_1 = N$. The condition of condensate 
existence, in the form of the gapless-spectrum requirement (14), gives
$\mu_1 = \mu_1(\mu_0,N)$ by the Hugenholtz-Pines relation [69]. These four 
equations uniquely define all four quantities: $N_0, N_1, \mu_0, \mu_1$ through
the parameters of the system in the stationary state.   

If one would assume the equivalence of the chemical potentials $\mu_0$ 
and $\mu_1$, this would overdefine the equations and would lead to inconsistences, 
as has been shown by Hohenberg and Martin [70]. Such inconsistences would lead 
either to the appearance of a gap in the spectrum or to the invalidity of the 
equation (16) for the condensate wave function. In turn, thermodynamic relations 
would break down and the poles of the single-particle and two-particle Green 
functions would be different, contradicting the results of Bogolubov [62] and 
Gavoret and Nozi\'{e}re [71], according to which the poles of these Green 
functions coincide, when the global gauge symmetry is broken. 

The Bogolubov canonical transformation
\be
\label{20}
\psi_1(\br,t) = \sum_k \; [ \; u_k(\br,t) b_k(t) +
v_k^*(\br,t) b_k^\dgr(t) \; ]
\ee
introduces the Bose operators $b_k(t)$ of collective excitations, {\it 
bogolons}, where $k$ is a quantum multi-index. This transformation makes 
it possible to diagonalize the grand Hamiltonian in the Hartree-Fock-Bogolubov
approximation. The coefficient functions $u_k$ and $v_k$ are found from this 
diagonalization which yields the Bogolubov equations that also give the 
Bogolubov spectrum $\varepsilon_k$. The bogolon number is not conserved and
bogolons cannot condense. This is evident from relation (20). Really, if the
bogolons would condense, resulting in nonzero $\langle b_k(t) \rangle$, then 
the average $\langle \psi_1({\bf r},t) \rangle$ would also become nonzero, 
which contradicts the basic definition (3).

Thus, atoms, whose total number $N$ is given, can form an equilibrium, 
as well as nonequilibrium, Bose-Einstein condensate. But the collective 
excitations, bogolons, whose number is not conserved, cannot be condensed. 
If in a nonequilibrium system one would notice the tendency of bogolons to
condense, this would simply imply the necessity of redefining the condensate
function of atoms in such a way that to preserve the principal definition 
(4), hence, avoiding the bogolon condensation. The same concerns other 
collective excitations that could arise in Bose-condensed systems, even if 
these excitations would have single-particle form [72]. The main point is 
that the collective excitations correspond to unconserved entities, so that 
they cannot condense. As is clear from the present section, bogolons, that 
are elementary collective excitations in a Bose-condensed system, by 
definition cannot condense whether in equilibrium or in any nonequilibrium 
case.

\section{Conserved versus unconserved particles}

As an explicit illustration demonstrating the basic difference between 
conserved and unconserved quasiparticles and the impossibility of 
equilibrium condensation for unconserved quasiparticles, let us consider 
boson quasiparticles with the effective spectrum 
\be
\label{21}
 \om_k = A k^n \;  ,
\ee
where $A$ and $n$ are positive parameters and $k$ is the modulus of a
$d$ - dimensional wave vector. This type of spectrum occurs in some problems 
of solid-state physics and some cosmological problems, where the matter with 
such a spectrum is named absolute stiff matter [73-76]. Quasiparticles are 
assumed to be in a large box of volume $V$, so that the sum over 
$d$-dimensional momenta can be replaced by the integral
$$
 \sum_k \; \longrightarrow \; 
V \int \frac{d\bk}{(2\pi)^d} \;
\longrightarrow \; 
V \int_0^\infty \frac{2 k^{d-1}dk}{(4\pi)^{d/2}\Gm(d/2)} \;  .
$$
The uniform density is denoted by $\rho \equiv N/V$.

\subsection{Equilibrium condensation of conserved quasiparticles}

Here we consider conserved quasiparticles, whose total number $N$ is 
fixed, therefore the density is constant, $\rho = const$. The critical 
temperature, where the condensation starts, is found in the usual 
way [18], yielding
\be
\label{22}
 T_c = A \left [ \; 
\frac{(4\pi)^{d/2}\Gm(d/2)n\rho}{2\Gm(d/n)\zeta(d/n)} 
\right ]^{n/d} \;  .
\ee
Taking into account that the gamma-function $\Gamma(x) > 0$ for $x > 0$ and
the Riemann zeta-function $\zeta(x) < 0$ in the interval $0 < x < 1$ tells us 
that there is no condensation for $d < n$. When $d = n$, then $T_c = 0$. And
$T_c$ is positive only for $d > n$. In the latter case, for the condensate 
fraction below $T_c$, one has 
\be
\label{23}
 n_0 \equiv \frac{N_0}{N} = 1 - 
\left ( \frac{T}{T_c} \right )^{d/n} \;  .
\ee

However, the formal existence of the critical temperature does not yet mean
that the condensate can really occur. In order to exist, it must represent
a stable statistical system. For this purpose, the system compressibility 
is to be finite, which is equivalent to the finiteness of the {\it stability 
ratio}
\be
\label{24}
 0 \; \leq  \; \frac{{\rm var}(\hat N)}{N} \; < \; \infty 
\ee
for all $N$, including thermodynamic limit [7,16,18,77]. Here
$$
 {\rm var}(\hat N) \equiv 
\lgl \hat N^2 \rgl - \lgl \hat N \rgl^2 \; , \qquad 
\hat N \equiv N_0 + \hat N_1 \;  .
$$
Calculating ratio (24), we use the fact that
$$
{\rm var}(\hat N) = {\rm var}(\hat N_1)  .
$$
Then, for $n < d/2$, we get
\be
\label{25}
 \frac{{\rm var}(\hat N)}{N}  = \frac{\zeta(d/n-1)}{\zeta(d/n)}
\left ( \frac{T}{T_c} \right )^{d/n} \qquad ( d > 2n) \;  .
\ee
In the interval $d/2 < n < d$, similarly to Ref. [78], we employ the 
generalized Bose functions, in whose definition the integrals are limited 
from below by the minimal wave number $k_{min} = (\rho / N)^{1/d}$. This 
gives
\be
\label{26}
 \frac{{\rm var}(\hat N)}{N}  \approx 
\frac{2(d-n)N^{(2n-d)/d}}{(4\pi)^{d/2} n (2n-d)\Gm(d/2)} 
\left ( \frac{T}{A\rho^{n/d}} \right )^2 \qquad ( n < d < 2n) \;  .
\ee

In particular cases, remembering that $T_c > 0$ exists only for $d>n$, we 
have the following. When $d=3$ and $n=2$, the 
critical temperature (22) is
\be
\label{27}
 T_c = 4\pi A \left [ \frac{\rho}{\zeta(3/2)} \right ]^{2/3} 
\qquad (d=3, \; n=2) \;  .
\ee
But ratio (26) diverges for large $N$,
\be
\label{28}
 \frac{{\rm var}(\hat N)}{N}  \approx 
\left ( \frac{T}{2\pi A \rho^{2/3}} \right )^2 N^{1/3} 
\qquad 
(d = 3, \; n = 2 ) \;  ,
\ee
hence, the stability condition (24) is not valid. 

When $d=2$ and $n=1$, the critical temperature is
\be
\label{29}
 T_c = 2A \sqrt{ \frac{3\rho}{\pi} } \qquad (d = 2, \; n=1 ) \;  .
\ee
But the stability ratio is strongly divergent, hence the system is unstable.

When $d=3$ and $n=1$, the critical temperature (22) becomes
\be
\label{30}
T_c  = A \left [ \frac{\pi^2\rho}{\zeta(3)} \right ]^{1/3} 
\qquad (d = 3, \; n=1 ) \;  ,
\ee
where we have used the value $\zeta(2) = \pi^2/6$. The stability ratio (25)
is finite:
\be
\label{31}
 \frac{{\rm var}(\hat N)}{N} = \frac{T^3}{6\rho A^3} 
\qquad (d = 3, \; n=1 ) \;  ,
\ee
telling us that the system is stable.

In this way, the uniform $d$-dimensional Bose gas of conserved quasiparticles 
with spectrum (21) can be Bose-condensed, provided that $d > 2n$.

\subsection{No equilibrium condensation of unconserved quasiparticles}

If the total number of quasiparticles is not fixed, but they can be created,
then, when decreasing temperature, no condensation occurs, but instead, the 
number of quasiparticles diminishes with temperature, which implies their 
decreasing density
\be
\label{32}
 \rho = \frac{2\Gm(d/n)\zeta(d/n)}{(4\pi)^{d/2}n\Gm(d/2)}
\left ( \frac{T}{A} \right )^{d/n} \;  .
\ee
This density is positive for $d>n$. In the particular cases, corresponding 
to those studied above, we have
\be
\label{33}
 \rho = \zeta\left ( \frac{3}{2} \right ) 
\left ( \frac{T}{4\pi A} \right )^{3/2} \qquad ( d = 3, \; n = 2) \; ,
\ee
\be
\label{34}
 \rho = \frac{\pi}{3} \left ( \frac{T}{2A} \right )^2 \qquad
 ( d = 2, \; n = 1) \; ,
\ee
\be
\label{35}
  \rho = \frac{\zeta(3)}{\pi^2} \left ( \frac{T}{A} \right )^3 \qquad
 ( d = 3, \; n = 1) \;  .
\ee
In all the cases, the density of unconserved quasiparticles diminishes with
temperature, tending to zero. But there is no condensation.

The similar situation happens in the case of equilibrium photons. In 
equilibrium, at lowering temperature, photons disappear instead of occupying 
the cavity ground state. But under an external pumping supporting photon total
number inside an optical microcavity, photons can condense [28]. Such a photon 
condensation, creating an ensemble of coherent photons is, actually, analogous 
to the functioning of lasers. 

When the lifetime of created quasiparticles is sufficiently long, a 
quasiequilibrium condensation of unconserved particles in a quasi-stationary 
state can be realized, as in the case of excitons [19-23,79]. But in absolute 
equilibrium, unconserved quasiparticles cannot condense.

\section{Vibrating optical lattices}

A very common quasiparticle, representing collective excitations, is phonon, 
characterizing density fluctuations caused by vibration of atoms in a lattice.
In addition to the standard solid-state physics, different lattices can be 
created by means of laser beams forming standing waves where cold atoms are 
caught [15,16,80,81]. Phonons are unconserved quasiparticles. Hence, we should 
expect that they cannot condense in equilibrium. However, in some cases one 
talks on condensation of phonons. Here the analysis is given explaining that 
the phonon condensation in equilibrium systems does not occur. The formal 
appearance of phonon condensation may happen in nonequilibrium situations, 
such as phase transitions. But defining phonons in a self-consistent way makes 
it possible to avoid phonon condensation in any system, equilibrium or not.

\subsection{Double-well optical lattices}

For generality, let us consider the case of atoms that can happen in two 
states of a lattice. A straightforward example of such two-state atoms are 
the atoms in double-well lattices [82-92] and different double-well potentials 
[93-95]. Another case could correspond to two-band lattices [96]. The usual 
single-band lattice can be considered as a particular case of a two-band 
lattice. 

The derivation of the lattice model starts with the standard Hamiltonian
$$
\hat H = \int \psi^\dgr(\br) H_1(\br) \psi(\br) \; d\br \; +
$$
\be
\label{36}
 + \; \frac{1}{2} 
\int \psi^\dgr(\br) \psi^\dgr(\br')\Phi(\br-\br') \psi(\br') \psi(\br) \; 
d\br d\br' \;  ,
\ee
where $\Phi({\bf r})$ is an atomic interaction potential and the single-atom
Hamiltonian is
\be
\label{37}
 H_1(\br) = -\; \frac{\nabla^2}{2m} + V_{lat}(\br) + 
U_{ext}(\br,t) \;  .
\ee
In the latter, $V_{lat}(\br)$ is a lattice potential and $U_{ext}(\br,t)$
is an external potential, such as the trapping potential. Generally, the 
external potential can be time-dependent. 

Atoms, located at the spatial points ${\bf r}_j$, are enumerated by the index 
$j = 1,2,\ldots,N$. Keeping in mind an insulating lattice, we can describe  
atomic states by the localized orbitals given by the Schr\"{o}dinger equation
\be
\label{38}
 \left [ -\; \frac{\nabla^2}{2m} + V_{lat}(\br) 
\right ] \psi_n(\br-\br_j) = E_{nj} \psi_n(\br-\br_j) \; .
\ee
The orbitals form an orthonormal basis with the property
$$
\int \psi_m^*(\br-\br_i) \psi_n(\br-\br_j)\; d\br = 
\dlt_{mn}\dlt_{ij} \;   .
$$  

The atomic field operator can be expanded over the orbital basis,
\be
\label{39}
 \psi(\br) = \sum_{nj} c_{nj} \psi_n(\br -\br_j) \;  .
\ee
We assume that at each location ${\bf r}_j$ there exists just one atom, which 
imposes the {\it unipolarity condition}
\be
\label{40}
\sum_n c_{nj}^\dgr c_{nj} = 1 \; , \qquad c_{mj} c_{nj} = 0 \;   .
\ee

The matrix element
\be
\label{41}
E_{ij}^{mn} = 
\int \psi_m^*(\br-\br_i) H_1(\br) \psi_n(\br-\br_j) \; d\br
\ee
of the single-atom Hamiltonian (37) is the sum 
\be
\label{42}
E_{ij}^{mn} = K_{ij}^{mn} + V_{ij}^{mn} + U_{ij}^{mn}
\ee
of the terms
$$
K_{ij}^{mn} \equiv \int \psi_m^*(\br-\br_i) 
\left ( -\; \frac{\nabla^2}{2m} \right ) \psi_n(\br-\br_j) \; d\br \; ,
$$
$$
V_{ij}^{mn} \equiv \int \psi_m^*(\br-\br_i) 
V_{lat}(\br) \psi_n(\br-\br_j) \; d\br \; ,
$$
$$
 U_{ij}^{mn} \equiv \int \psi_m^*(\br-\br_i) 
U_{ext}(\br,t) \psi_n(\br-\br_j) \; d\br \; .
$$
And the eigenvalues of Eq. (38) are
\be
\label{43}
 E_{nj} = K_{jj}^{nn} + V_{jj}^{nn} \;  .
\ee
Then the matrix elements (42) can be written as
\be
\label{44}
E_{ij}^{mn} = \dlt_{mn} \dlt_{ij} E_{nj} + U_{ij}^{mn} \; .
\ee

We assume that only two lowest bands are occupied, so that $n = 1,2$. Then 
it is possible to introduce the pseudospin operators
$$
S_j^x = \frac{1}{2} \left ( c_{1j}^\dgr c_{1j} - 
c_{2j}^\dgr c_{2j}  \right ) \; , \qquad 
S_j^y = \frac{i}{2} \left ( c_{1j}^\dgr c_{2j} - 
c_{2j}^\dgr c_{1j} \right ) \; ,
$$
\be
\label{45}
 S_j^z = \frac{1}{2} \left ( c_{1j}^\dgr c_{2j} + 
c_{2j}^\dgr c_{1j} \right ) \; ,  
\ee
which gives the atomic operators 
$$
c_{1j}^\dgr c_{1j} = \frac{1}{2} + S_j^x \; , \qquad
c_{2j}^\dgr c_{2j} = \frac{1}{2} - S_j^x \; ,
$$
\be
\label{46}
 c_{1j}^\dgr c_{2j} = S_j^z - i S_j^y\; , \qquad
c_{2j}^\dgr c_{1j} = S_j^z + i S_j^y \;   .
\ee
The pseudospin operators satisfy the standard spin algebra. And the atomic 
operators can satisfy either boson or fermion commutation relations.

To better understand the physical meaning of the pseudospin operators (45),
we can introduce the left and right location operators
\be
\label{47}
 c_{jL} = \frac{1}{\sqrt{2} } \; ( c_{1j} + c_{2j} ) \;  , \qquad
c_{jR} = \frac{1}{\sqrt{2} } \; ( c_{1j} - c_{2j} )
\ee
describing the location of atoms in the left or right wells of a double well
that is centered at the lattice point ${\bf r}_j$. Then the pseudospin 
operators (45) become
$$
S_j^x = \frac{1}{2} \left ( c_{jL}^\dgr c_{jR} + 
c_{jR}^\dgr c_{jL} \right ) \; , \qquad
S_j^y = -\; \frac{i}{2} \left ( c_{jL}^\dgr c_{jR} - 
c_{jR}^\dgr c_{jL} \right ) \; ,
$$
\be
\label{48}
S_j^z = \frac{1}{2} \left ( c_{jL}^\dgr c_{jL} - 
c_{jR}^\dgr c_{jR} \right ) \; .
\ee
This shows that $S_j^x$ corresponds to tunneling transitions between 
the left and right locations, $S_j^y$ characterizes the Josephson current 
between these locations, and $S_j^z$ describes the population imbalance 
between the wells. Thus, the operator $S_j^z$ can be called the 
{\it deformation operator}, since, when atoms move to one of the wells, 
the spatial system symmetry is broken and the system becomes deformed. For 
a finite system, such a deformation leads to the change of its shape.     

Substituting expansion (39) into Hamiltonian (36), we denote the matrix 
elements of the interaction potential as $A(r_{ij})$, $B(r_{ij})$, and 
$C(r_{ij})$, whose detailed definition can be found in Refs. [89,90,97] 
and where
$$
 r_{ij} \equiv |\; \br_{ij} \; | \; , \qquad 
\br_{ij} \equiv \br_i - \br_j \;  .
$$
Also, we introduce the following notation:
$$
E_0 \equiv \frac{1}{2N} \sum_j \left ( V_{jj}^{11} +
V_{jj}^{22} \right ) \; , \qquad
p_j^2 \equiv m \left ( K_{jj}^{11} + K_{jj}^{22} \right ) \; ,
$$
\be
\label{49}
\Om_j \equiv E_{jj}^{22} - E_{jj}^{11} + \sum_i C(r_{ij}) \; , 
\qquad
h_j \equiv E_{jj}^{12} + E_{jj}^{21} = 
U_{jj}^{12} + U_{jj}^{21} \; .
\ee
In this way, Hamiltonian (36) transforms into
$$
\hat H = NE_0 + \sum_j \left ( \frac{p_j^2}{2m} \; - \; 
\Om_j S_j^x + h_j S_j^z \right ) \; +
$$
\be
\label{50}
+ \; \sum_{i\neq j} \left [ \frac{1}{2} \; A(r_{ij}) +
B(r_{ij}) S_i^x S_j^x - I(r_{ij}) S_i^z S_j^z \right ] \;   .
\ee
Note that the expression $h_j$, playing the role of an external field, 
is nonzero only when the external potential $U_{ext}({\bf r},t)$ is not 
uniform. As soon as $U_{ext}$ is uniform in space, the effective field 
$h_j$ is zero.

\subsection{Operators of atomic deviations}

Taking into account vibrational degrees of freedom can be done by introducing 
the operators of atomic deviations ${\bf u}_j$ by the equation
\be
\label{51}
 \br_j = \ba_j + \bu_j \;  ,
\ee
where ${\bf a}_j$ is a fixed vector related to a $j$-th lattice site. Since the
interaction terms in Eq. (50) depend on the difference ${\bf r}_{ij}$, it is 
convenient to use the relative deviation
\be
\label{52}
 \bu_{ij} \equiv \bu_i - \bu_j \;  .
\ee

Treating ${\bf u}_j$ as small deviations around ${\bf a}_j$, we expand the 
interaction terms in Hamiltonian (50) in powers of ${\bf u}_{ij}$. In this
expansion, we use the notations
$$
A_{ij} \equiv A(a_{ij}) \; , \qquad 
A_{ij}^\al \equiv \frac{\prt A_{ij}}{\prt a_i^\al} \; , 
\qquad 
A_{ij}^{\al\bt} \equiv 
\frac{\prt^2 A_{ij}}{\prt a_i^\al \prt a_j^\bt} \; ,
$$
where
$$
 a_{ij} \equiv | \ba_{ij} | \; , \qquad 
\ba_{ij} \equiv \ba_i - \ba_j \;  .
$$
Analogous notations are employed in the expansions of other interaction 
terms.
    
Since the expression
$$
A \equiv \frac{1}{N} \sum_{ij} A_{ij} = \sum_j A_{ij}
$$
does not depend on the atomic indices, we have
$$
\sum_j A_{ij}^\al = \frac{\prt A}{\prt a_i^\al} = 0 \; , 
\qquad
\sum_j A_{ij}^{\al\bt} = -\;
\frac{\prt^2 A}{\prt a_i^\al\prt a_i^\bt} = 0 \; ,   
$$
$$
\sum_{ij} A_{ij}^\al u_{ij}^\al = 
2 \sum_i u_i^\al \; \sum_j A_{ij}^\al = 0 \; .
$$
In the above summations, we keep in mind that $i \neq j$, though this is 
not shown explicitly for the brevity of notation. Alternatively, we may 
set $A_{ii} \equiv 0, I_{ii} \equiv 0, B_{ii} \equiv 0$. 

Following this way [97,98], we come to the Hamiltonian
$$
\hat H = \left ( E_0 + \frac{A}{2} \right ) N + \sum_j \left (
\frac{p_j^2}{2m} \; - \; \Om_j S_j^x + h_j S_j^z \right ) +
$$
$$
+ \sum_{ij} \; \sum_{\al\bt} \left [ -\;
\frac{1}{4} \; A_{ij}^{\al\bt} u_{ij}^\al u_{ij}^\bt +  
\left ( B_{ij} + \sum_\al B_{ij}^\al u_{ij}^\al \; - \;
\frac{1}{2} \sum_{\al\bt}  B_{ij}^{\al\bt} u_{ij}^\al u_{ij}^\bt 
\right ) S_i^x S_j^x  - \right.
$$
\be
\label{53}
 - \left. \left ( I_{ij} + \sum_\al I_{ij}^\al u_{ij}^\al \; - \;
\frac{1}{2} \sum_{\al\bt} I_{ij}^{\al\bt} u_{ij}^\al u_{ij}^\bt 
\right ) S_i^z S_j^z \right ] \;  .
\ee

Vibrational and atomic degrees of freedom in the four-operator terms can be 
decoupled by the relation
\be
\label{54}
  u_{ij}^\al u_{ij}^\bt S_i^\gm S_j^\gm =
\lgl  u_{ij}^\al u_{ij}^\bt \rgl S_i^\gm S_j ^\gm +
u_{ij}^\al u_{ij}^\bt \lgl S_i^\gm S_j^\gm \rgl - 
\lgl u_{ij}^\al u_{ij}^\bt \rgl \lgl  S_i^\gm S_j ^\gm \rgl \; ,
\ee
while keeping untouched the lower-order terms. We suppose that the pseudospin 
correlation functions $\langle S_i^\alpha S_j^\alpha \rangle$ weakly depend 
on the lattice indices, so that 
$$
 \sum_{j(\neq i)} B_{ij}^{\al\bt} \lgl S_i^x S_j^x \rgl \cong -\;
\frac{\prt^2}{\prt a_i^\al\prt a_i^\bt} \; 
\sum_{j(\neq i)} B_{ij} \lgl S_i^x S_j^x \rgl \cong 0 \; ,
$$
$$
\sum_{j(\neq i)} I_{ij}^{\al\bt} \lgl S_i^z S_j^z \rgl \cong -\;
\frac{\prt^2}{\prt a_i^\al\prt a_i^\bt} \; 
\sum_{j(\neq i)} I_{ij} \lgl S_i^z S_j^z \rgl \cong 0 \; .
$$
Thus we come to the Hamiltonian
\be
\label{55}
 \hat H = \widetilde E_0 N + H_{ph} + H_{at} +
H_{int} \;  .
\ee
Here, in the first term
\be
\label{56}
 \widetilde E_0 \equiv E_0 + \frac{A}{2} + \frac{1}{N} \sum_{ij}
\sum_{\al\bt} \left ( I_{ij}^{\al\bt} \lgl S_i^z S_j^z \rgl -
B_{ij}^{\al\bt} \lgl S_i^x S_j^x \rgl \right ) 
\lgl u_i^\al u_j^\bt - u_j^\al u_j^\bt  \rgl \;  .
\ee
The second term is the effective Hamiltonian of dressed phonons
\be
\label{57}
 H_{ph} = \sum_j \frac{p_j^2}{2m} + 
\frac{1}{2} \sum_{ij} \sum_{\al\bt} 
\Phi_{ij}^{\al\bt} u_i^\al u_j^\bt \; ,
\ee
with the renormalized dynamic matrix
\be
\label{58}
 \Phi_{ij}^{\al\bt} \equiv A_{ij}^{\al\bt} - 
2 I_{ij}^{\al\bt} \lgl S_i^z S_j^z \rgl +
2 B_{ij}^{\al\bt} \lgl S_i^x S_j^x \rgl \;  .
\ee
The third term is the Hamiltonian of dressed atoms
\be
\label{59}
 H_{at} = \sum_j \left ( h_j S_j^z - \Om_j S_j^x \right ) +
\sum_{ij} \left ( \widetilde B_{ij} S_i^x S_j^x - 
\widetilde I_{ij} S_i^z S_j^z \right ) \;  ,
\ee
with the renormalized atomic interactions
\be
\label{60}
 \widetilde B_{ij} \equiv B_{ij} + 
\sum_{\al\bt} B_{ij}^{\al\bt} \lgl u_i^\al u_j^\bt - 
u_j^\al u_j^\bt \rgl \; , \qquad
 \widetilde I_{ij} \equiv I_{ij} + 
\sum_{\al\bt} I_{ij}^{\al\bt} \lgl u_i^\al u_j^\bt - 
u_j^\al u_j^\bt \rgl \; .
\ee
And the fourth term describes atom-phonon interactions,
\be
\label{61}
  H_{int} = - 2 \sum_i \sum_\al F_i^\al u_i^\al \; ,
\ee
in which the effective force acting on an atom is
\be
\label{62}
 F_i^\al \equiv \sum_j F_{ij}^\al \;  ,
\ee
where
\be
\label{63}
 F_{ij}^\al \equiv I_{ij}^\al S_i^z S_j^z - 
B_{ij}^\al S_i^x S_j^x \;  .
\ee

\subsection{Pseudophonon and phonon operators}

In order to introduce phonon operators, let us define in the standard way
the phonon frequency $\omega_{ks}$ by the eigenproblem
\be
\label{64}
 \frac{1}{m} \sum_j \sum_\bt 
\Phi_{ij}^{\al\bt} e^{i\bk\cdot\ba_{ij}} e_{ks}^\bt  = 
\om_{ks}^2 e_{ks}^\al \; ,
\ee
where ${\bf k}$ is momentum, $s = 1,2,3$ is polarization index, and 
${\bf e}_{ks}$ is a polarization vector. 

If one defines the phonon operators $d_{ks}$ by the usual relations
$$
\bp_j = - \; \frac{i}{\sqrt{2N}} \; \sum_{ks} \; \sqrt{m \om_{ks} } \;
\bbe_{ks} \left ( d_{ks} - d_{-ks}^\dgr \right ) e^{i\bk\cdot\ba_j} \; ,
$$
\be
\label{65}
  \bu_j = \frac{1}{\sqrt{2N}} \; 
\sum_{ks} \frac{\bbe_{ks}}{ \sqrt{m\om_{ks}} }
\left ( d_{ks} + d_{-ks}^\dgr \right ) e^{i\bk\cdot\ba_j} \; ,
\ee
the total Hamiltonian (55) does not become diagonal with respect to these
operators $d_{ks}$. That is, such operators are not true phonon operators,
but rather some {\it pseudophonon operators}. To diagonalize the Hamiltonian,
we need to accomplish an additional canonical transformation
\be
\label{66}
 d_{ks} = b_{ks} + \frac{2}{\sqrt{2N}\; \om_{ks}\sqrt{m\om_{ks}} } \;
\sum_i \sum_\al F_i^\al e_{ks}^\al e^{-i\bk\cdot\ba_i} \;  .
\ee
This is equivalent [97] to introducing the phonon operators $b_{ks}$ by the 
nonuniform relations
$$
\bp_j = - \; \frac{i}{\sqrt{2N}} \; \sum_{ks} \; \sqrt{m \om_{ks} } \;
\bbe_{ks} \left ( b_{ks} - b_{-ks}^\dgr \right ) e^{i\bk\cdot\ba_j} \; ,
$$
\be
\label{67}
 \bu_j = {\bf\Dlt}_j + \frac{1}{\sqrt{2N}} \; \sum_{ks} \;
 \frac{\bbe_{ks}}{ \sqrt{m\om_{ks}} } \left ( b_{ks} + b_{-ks}^\dgr 
\right ) e^{i\bk\cdot\ba_j} \;,
\ee
where the additional vector term has the components
\be
\label{68}
 \Dlt_i^\al = 2 \sum_j \sum_\bt \gm_{ij}^{\al\bt} F_j^\bt \;  ,
\ee
with the matrix
\be
\label{69}
 \gm_{ij}^{\al\bt} \equiv \frac{1}{N} \sum_{ks} \; 
\frac{e_{ks}^\al e_{ks}^\bt}{m\om_{ks}^2} \; 
e^{i\bk\cdot\ba_{ij}} \;  .
\ee

As a result of this nonuniform transformation (67), Hamiltonian (55) 
takes the form
\be
\label{70}
 \hat H = N \widetilde E_0 +  \widetilde H_{ph} + 
H_{at} + H_4 \; ,
\ee
with the diagonal phonon part
\be
\label{71}
 \widetilde H_{ph} = \sum_{ks} \om_{ks} \left ( b_{ks}^\dgr b_{ks} 
+ \frac{1}{2} \right ) \;  ,
\ee
and with an additional four-atom interaction term
\be
\label{72}
H_4 = - \; \frac{4}{N} \; \sum_{ij} \sum_{fg} \sum_{\al\bt}
F_{ij}^\al \gm_{jf}^{\al\bt} F_{fg}^\bt \;   .
\ee

Since the Hamiltonian is diagonal in operators $b_{ks}$, it is these 
operators that should be accepted as real phonon operators.

\subsection{Shape distortion operator}

Operator (68), entering transformation (67), characterizes the shift
of the deviation operator ${\bf u}_j$ caused by the existence of atomic
degrees of freedom described by the pseudospin operators $S_j^\alpha$. 
Such a shift of atoms from their initial positions ${\bf a}_j$ would lead 
to the distortion of the considered sample, because of which the operator
${\bf \Delta}_j$ can be called the {\it distortion operator}. In the case
of a finite system, this deformation results in the change of the sample 
shape. To illustrate more explicitly the form of the distortion operator,
one needs to specify the phonon frequency.

If we take the phonon frequency in the Einstein approximation
\be
\label{73}
 \om_{ks} = \om_0 \;  ,
\ee      
then matrix (69) is
$$
 \gm_{ij}^{\al\bt} = 
\frac{\dlt_{\al\bt}\dlt_{ij} }{m\om_0^2} \;  .
$$
This gives the distortion operator
\be
\label{74}
\Dlt_i^\al = \frac{2}{m\om_0^2}\; F_i^\al \; .
\ee

Another common case is the Debye approximation for the frequency
\be
\label{75}
 \om_{ks} = \Theta(k_D - k) ck \;  ,
\ee
in which $\Theta(k)$ is a unit step function and 
$$
k_D \equiv \left ( 6\pi^2\rho \right )^{1/3}
$$
is the Debye wave vector modulus. Then for matrix (69), we get
$$
 \gm_{ij}^{\al\bt} = \frac{\dlt_{\al\bt}}{2\pi^2\rho mc^2} \;
\frac{{\rm Si}(k_D a_{ij})}{a_{ij}} \;  ,
$$
where ${\rm Si}(x)$ is the sine integral
$$
{\rm Si}(x) \equiv \int_0^x \frac{\sin t}{t} \; dt \;   .
$$
The latter possesses the asymptotic properties
$$
 {\rm Si}(x) \simeq x \; - \; \frac{x^3}{18} \qquad (x\ra 0) \;  ,
$$
$$
{\rm Si}(x) \simeq \frac{\pi}{2} \; - \; \frac{\cos x}{x} \qquad
(x\ra \infty ) \; .
$$
The distortion operator becomes
\be 
\label{76}
 \Dlt_i^\al = \frac{1}{\pi^2\rho mc^2} \; \sum_j F_j^\al \;
\frac{{\rm Si}(k_D a_{ij})}{a_{ij} } \;  .
\ee

\subsection{No equilibrium phonon condensation}

Real phonons are not conserved quasiparticles and, as is seen from the phonon 
Hamiltonian (71), 
\be
\label{77}
 \lgl b_{ks} \rgl = 0 \;  ,
\ee
that is, equilibrium phonons cannot condense. At the same time, as follows 
from Eq. (67), the average deviation can be nonzero, since
\be
\label{78}
 \lgl \bu_j \rgl = \lgl {\bf\Dlt}_j \rgl \;  .
\ee

To explicitly show how the deviation can arise, let us resort to the mean-field 
approximation for the correlation function
\be
\label{79}
  \lgl S_i^\al S_j^\al \rgl = \lgl S_i^\al \rgl \lgl S_j^\al \rgl
\qquad ( i \neq j ) \; .
\ee
Denoting the mean pseudospin
\be
\label{80}     
C_J^\al \equiv \lgl S_j^\al \rgl \;  ,
\ee
we have for the average force (62)
\be
\label{81}
 \lgl F_i^\al \rgl = 
\sum_j \left ( I_{ij}^\al C_i^z C_j^z - B_{ij}^\al C_i^x C_j^x 
\right ) \; .
\ee
Therefore, the observed distortion (68) is, generally, nonzero. For instance,
taking the phonon frequency in the Einstein approximation yields
\be
\label{82}
  \lgl \Dlt_i^\al \rgl = \frac{2}{m\om_0^2} \; \sum_j \left (
I_{ij}^\al C_i^z C_j^z - B_{ij}^\al C_i^x C_j^x \right ) \; .
\ee
When $C_j^\alpha$ is zero itself or does not depend on the index $j$, then 
distortion (82) is zero. But for $C_j^\alpha \neq 0$ and depending on $j$, 
there appears a nonzero distortion. For double wells, the average $C_j^z$
plays the role of an order parameter distinguishing a disordered phase with
$C_j^z = 0$ from an ordered phase where $C_j^z \neq 0$. The ordered phase
exists at low temperatures below a critical temperature [89,90]. The 
ordering means the appearance of atomic imbalance between the wells of a 
double well. The order-disorder phase transition occurs if there is no 
external force $h_j$ defined in Eq. (49). When the latter is absent, the
system Hamiltonian (55) is invariant under the inversion $S_j^z \ra - S_j^z$,
which implies the existence of a degenerate ground state. This situation 
is analogous to the Jahn-Teller effect [99,100], where the existence of a 
degenerate ground state of a molecule leads to the distortion of the latter. 

The appearance of a nonzero distortion (82) could be associated with the 
condensation of pseudophonons described by the operators $d_{ks}$. However,
it is evident that this condensation simply means the shift of the average 
location of each atom occurring at the point of the order-disorder phase 
transition. At this point the system is unstable. After the phase transition 
has happened, one has to change the definition of atomic locations and, 
respectively, the definition of the deviation from this location. 

Alternatively, it is possible to define the atomic locations as the averages
\be
\label{83}     
 \ba_j \equiv \lgl \br_j \rgl \;  .
\ee
Then, in view of relation (51), one always has zero deviation
\be
\label{84}
 \lgl \bu_j \rgl = 0 \;  ,
\ee
and, because of equality (78), zero distortion
\be
\label{85}
  \lgl {\bf\Dlt}_j \rgl = 0 \; .
\ee
In this case, the occurrence of the order-disorder phase transition is 
exhibited in the sharp variation, e.g., with temperature, of the average
atomic location (83). But there is neither phonon nor pseudophonon 
condensation.    

Condition (84) can be preserved in any given approximation by defining the
grand Hamiltonian
$$
 H = \hat H - \sum_j \vec{\lbd}_j \cdot \bu_j \;  ,
$$
in which the Lagrange multipliers $\lambda_j$ are chosen so that to cancel
the linear in the deviations ${\bf u}_j$ terms, thus, guaranteeing the 
statistical condition (84).

Even more, definition (83) can be extended to nonequilibrium systems by 
introducing atomic deviations through the equation
$$
\br_j(t) = \ba_j(t) + \bu_j(t) \;   ,
$$
in which the first term is the average
$$
  \ba_j(t) \equiv \lgl \br_j(t) \rgl \; .
$$
I that case, by definition, 
$$
 \lgl \bu_j(t) \rgl = 0 \;  ,
$$
hence there is no symmetry breaking:
$$
 \lgl b_{ks}(t) \rgl = 0 \;  .
$$
Defining in such a self-consistent way atomic deviations makes phonon 
condensation absent in any nonequilibrium system.

\section{No equilibrium magnon condensation}

Collective excitations, characterizing spin oscillations, or spin waves, 
are magnons. Their definition can be done by the Holstein-Primakoff
transformation [101,102] for the ladder spin operators
$$
 S_j^\pm \equiv S_j^x \pm i S_j^y \;  .
$$
This transformation for the spin operator ${\bf S}_j$, located at 
${\bf r}_j$, reads as
$$
  S_j^+ = \sqrt{2S - b_j^\dgr b_j } \; b_j \; , \qquad
 S_j^- = b_j^\dgr \; \sqrt{2S - b_j^\dgr b_j } \; ,
$$
\be
\label{86} 
S_j^z = S - b_j^\dgr b_j \; ,
\ee
where $S$ is the spin value and $b_j$ are Bose operators representing magnons. 
The square root of operators is to be understood as Taylor series in powers
of the magnon density operator
$$
 \hat n_j \equiv b_j^\dgr b_j \;  ,
$$
which implies the assumption on the small deviation $S - S_j^z$. The 
eigenvalues of the magnon-density operator $\hat{n}_j$ are $0,1,2,\ldots,2S$.
In the lowest approximation,
\be
\label{87}
  S_j^+ \simeq \sqrt{2S} \; b_j \; , \qquad
S_j^- \simeq \sqrt{2S} \; b_j^+ \; .
\ee

The Holstein-Primakoff magnons, by definition, characterize small spin 
fluctuations around the ground state, for which the average spin is directed 
along the $z-$axis,
\be
\label{88}
\lgl \bS_j \rgl = S \bbe_z \; ,
\ee
so that there are no transverse components:
\be
\label{89}
\lgl S_j^x \rgl = \lgl S_j^y \rgl = 0 \; .
\ee
The latter means that
\be
\label{90}
\lgl b_j \rgl = 0 \; .
\ee

By construction, the Holstein-Primakoff representation (86) presupposes 
that magnons describe small fluctuations around the saturated magnetization
directed along the $z$-axis [101]. The appearance of magnon condensation 
would mean the rotation of the magnetization because of the arising nonzero
transverse magnetization. But then the total magnetization is not directed 
anymore along the $z$-axis, as is assumed in the Holstein-Primakoff 
representation (86). That is, the latter is not applicable, so that magnons 
are not well defined.     

If in calculations one formally meets magnon condensation in an
equilibrium state, when $\lgl b_j\rgl$ becomes nonzero, this implies that
at that point a phase transition occurs, with the rotation of the average
magnetization. Then one has to redefine the system ground state, for which
the Holstein-Primakoff transformation (86) would be valid, since  
representation (86) is not applicable if magnons would be condensed. The
rotation of system magnetization does not mean magnon condensation, but 
means the change of the system ground state.

Magnons are not conserved and in equilibrium they cannot condense,
$\lgl b_j \rgl =0$, similarly to the impossibility of equilibrium phonon
condensation or to the equilibrium condensation of other unconserved
quasiparticles. Impossibility of the equilibrium magnon condensation has 
been stressed by Mills [36].     

There are interpretations of some magnetic effects in nonequilibrium systems
as being due to magnon condensation. These can be the systems that are 
subject to a sufficiently strong external alternating field, whose frequency 
plays the role of a magnon chemical potential determining the nonequilibrium 
density of magnons [103-108]. Or these can be the spin systems prepared in a 
strongly nonequilibrium initial state and relaxing to equilibrium by 
essentially nonlinear dynamics [109-120]. The magnon condensation is 
manifested as the emerging state of precessing spins, where all spins precess 
with the same frequency and phase. This nonequilibrium magnon condensation is 
accompanied by the arising off-diagonal order, when the averages
$$
\lgl S_j^\pm \rgl \; \propto \; e^{i\om_0 t}
$$
are nonzero, where $\omega_0$ is the Zeeman frequency. The nonequilibrium 
magnon condensation is an example of {\it dynamic coherence}, contrary
to {\it static coherence} related to the Bose-Einstein condensation of 
conserved atoms. 

As is mentioned in the Introduction, there exists a number of articles 
claiming the occurrence of magnon condensation in equilibrium stable magnets. 
Such claims are based on confusion, when either the Holstein-Primakoff 
representation is used outside of its region of validity or transformations 
are employed to some auxiliary quasiparticles that have nothing to do with 
magnons. One could talk on magnon condensation at magnetic phase transitions 
[121-123], which does not contradict the above conclusion on the impossibility 
of equilibrium magnon condensation, since the point of a phase transition is 
that where the system loses its stability, hence, an equilibrium system does 
not exist. But the occurrence of magnon condensation in the whole range of
thermodynamic parameters, in an equilibrium system, is principally impossible.

\section{Operators of auxiliary quasiparticles}

I many cases, one accomplishes operator canonical transformations aiming at
simplifying the solution of a problem. In such transformations, the newly
introduced operators correspond to auxiliary quiasiparticles that do not 
necessarily represent physical particles, but can be just mathematical tools, 
not related to any physical objects. In many cases the statistics of such 
auxiliary quasiparticles are even not uniquely defined, and the latter can 
be equally treated as bosons or as fermions. Depending on whether the auxiliary 
quasiparticles are conserved and prescribed to be bosons, they can formally 
exhibit condensation in equilibium systems. In this and in the following 
section, some examples of auxiliary quasiparticles are given.

\subsection{Jordan-Wigner transformation}

As an example of a trasformation, introducing auxiliary quasiparticle, let us 
mention the mapping of spin operators of spin $1/2$, for a one-dimensional 
chain, onto the fermionic operators. After this mapping, one expresses the 
fermionic operators in terms of bosonic ones using Abelian bosonization [124,125].
   
One considers the spin operators
\be
\label{91}
\bS_j = \frac{1}{2} \; \vec{\sgm}_j \;   ,
\ee
where $\vec{\sigma}_j$ is the vector Pauli matrix. The Jordan-Wigner [126] 
transformation 
\be
\label{92}    
S_j^+ = \exp \left ( i\pi \sum_{n=1}^{j-1} a_n^\dgr a_n 
\right ) a_j^\dgr \; , \quad
S_j^- = \exp \left ( - i\pi \sum_{n=1}^{j-1} a_n^\dgr a_n 
\right ) a_j \; , \quad S_j^z = a_j^\dgr a_j \; - \; \frac{1}{2} \; ,
\ee
maps the ladder spin operators onto fermionic operators $a_j$. The 
latter do not correspond to some really existing particles or quasiparticles, 
but they just allow one to solve the studied one-dimensional problem.

One may notice that
\be
\label{93}
S_j^+ S_j^- = a_j^\dgr a_j \;   .
\ee
Thence, the last of transformations (92) can be written in the form
$$
 S_j^z = S_j^+ S_j^- \; - \; \frac{1}{2} \;  .
$$
This is in agreement with the general properties of $S$-spin operators
$$
\bS_i \cdot \bS_j = \frac{1}{2} \left ( S_i^+ S_j^- + S_j^+ S_i^- 
\right ) + S_i^z S_j^z - \dlt_{ij} S_j^z \; , 
$$
$$
 \bS_j^2 = S_j^+ S_j^- + ( S_j^z )^2 - S_j^z = S ( S + 1 ) \; .
$$

The inverse transformation reads as
\be
\label{94}
 a_j^+ = \exp \left ( -i \pi \sum_{n=1}^{j-1} S_n^+ S_n^- 
\right ) S_j^+ \; , \qquad  
a_j^- = \exp \left ( i \pi \sum_{n=1}^{j-1} S_n^+ S_n^- 
\right ) S_j^- \;.
\ee
The total number of the auxiliary quasiparticles is not conserved, since
\be
\label{95}
 \sum_j \lgl a_j^\dgr a_j \rgl = 
\frac{N}{2} + \sum_j \lgl S_j^z \rgl \;  .
\ee

In this case, the auxiliary fermionic operators do not represent physical 
particles. If in the process of some canonical transformation one would
employ a mapping to auxiliary bosonic operators, one could talk on the 
possible condensation of the related auxiliary quasiparticles. Recall, 
however, that equilibrium condensation can happen only for conserved 
quasiparticles.

\subsection{Condensation of Schwinger bosons}

Auxiliary quasiparticles can be introduced, for example, by means of 
the transformation
\be
\label{96}
 \bS_j = \frac{1}{2} 
\sum_{\al\bt} a_{j\al}^\dgr \vec{\sgm}_{\al\bt} a_{j\bt}  
\ee
mapping the spin operator ${\bf S}_j$ at the spatial location $\br_j$
to quasiparticle operators $a_{j \alpha}$, in which $\alpha = 1,2$ 
enumerates two admissible states, for instance, spin projections. The
operator
\be
\label{97}
 \vec{\sgm}_{\al\bt} = \sum_\gm \sgm_{\al\bt}^\gm \bbe_\gm \qquad
(\gm = x,y,z)   
\ee
is a linear combination of the elements $\sigma^\gamma_{\alpha \beta}$
of the Pauli matrices, where ${\bf e}_\gamma$ is a unit vector in the 
direction labelled by $\gamma = x,y,z$. Thus, the components of 
vector (96) are
$$
S_j^\gm = \frac{1}{2} 
\sum_{\al\bt} a_{j\al}^\dgr \sgm_{\al\bt}^\gm a_{j\bt} \; .
$$
Explicitly, this transformation takes the form
\be
\label{98}
 S_j^+ = a_{j1}^\dgr a_{j2} \; , \qquad S_j^- = a_{j2}^\dgr a_{j1} \; ,
\qquad S_j^z = \frac{1}{2} 
\left ( a_{j1}^\dgr a_{j1} - a_{j2}^\dgr a_{j2} \right ) \;  .
\ee
The quasiparticle operators satisfy the constraint
\be
\label{99}
 \sum_\al a_{j\al}^\dgr a_{j\al} = 2S \; .
\ee

It is important to emphasize that $a_{j \alpha}$ can be either bosonic 
or fermionic operators, thus, representing any type of quasiparticles,
whether bosons or fermions. Transformation (98) was, first, introduced by
Bogolubov [62,127], with the operators $a_{j \alpha}$ representing electrons,
that is, fermions. Schwinger [128] treated $a_{j \alpha}$ as representing
bosons, because of which the latter are called the Schwinger bosons [129].
If the index $\alpha$ is interpreted as spin projection, then the Schwinger 
bosons are fictitious quasiparticles that cannot exist in reality, which 
follows from their unphysical nature, as far as they are bosons possessing 
spin one-half. But they can be real particles when the index $\alpha$ 
corresponds to the enumeration of some other components. Also, pairons 
[130], that are tightly bound pairs of particles, could be connected with 
Schwinger bosons [131]. 

In view of constraint (99), the total number of quasiparticles in conserved:
\be
\label{100}
 \sum_j \sum_\al \lgl a_{j\al}^\dgr a_{j\al} \rgl = 2 S N \;  .
\ee
Being treated as bosons, conserved quasiparticles can condense, which has 
been used in the interpretation of magnetic transitions [132,133], though 
interpreting the magnetic order-disorder phase transition as the Bose-Einstein 
condensation of Schwinger bosons can lead to incorrect first-order 
transition, instead of the correct second-order one [134].

Schwinger bosons are conserved quasiparticles and are principally different
from the unconserved Holstein-Primakoff magnons [134]. This is why the 
former can condense in equilibrium systems, while the latter cannot.

\subsection{Auxiliary slave bosons}

In the theory of strongly correlated Fermi systems, one uses the 
slave-particle approach based on the factorization of the electron field 
operator into two operators, one of which is bosonic and another, fermionic.
There are two alternatives that are termed slave boson and slave fermion
representations.

In the {\it slave boson representation} [135-137], the electron operator 
$c_{i \sigma}$, describing the annihilation of an electron at the point 
${\bf r}_i$ with spin $\sigma$, is formally split into the product
\be
\label{101}
 c_{i\sgm} = b_i^\dgr f_{i\sgm} \;  ,
\ee
where $b_i$ is a spinless bosonic operator and $f_{i \sigma}$ is a fermionic 
operator with spin $\sigma$. The operators satisfy the no-double-occupancy 
constraint
\be
\label{102}
 b_i^\dgr b_i + \sum_\sgm f_{i\sgm}^\dgr f_{i\sgm} = 1 \; .
\ee
The bosonic operator $b_i$ is said to represent an auxiliary charged spinless 
quasiparticle, named {\it holon}, while the fermionic operator $f_{i \sigma}$, 
an auxiliary neutral spin one-half quasiparticle {\it spinon}.

By this definition, it is clear that the holons and spinons are fictitious 
quasiparticles. According to constraint (102), the total number of these 
auxiliary quasiparticles is conserved:
\be
\label{103}
 \sum_i \left ( \lgl b_i^\dgr b_i \rgl + 
\sum_\sgm \lgl f_{i\sgm}^\dgr f_{i\sgm} \rgl \right ) = N \;  ,
\ee
where $N$ is the total number of electrons. The number of holons separately, 
strictly speaking, is not conserved. But it can be treated as quasiconserved,
when the number of spinons is limited by a value smaller than $N$. If the 
number of spinons is fixed by the doping concentration, as is usually done, 
then holons become conserved exactly. Therefore, it is admissible to talk 
about holon condensation [138] or holon pair condensation [139].

In the {\it slave fermion representation} [140-142], one formally splits the 
electron operator into the product
\be
\label{104}
  c_{i\sgm} = f_i^+ b_{i\sgm} \; ,
\ee
with a fermionic operator $f_i$ corresponding to spinless charged fermionic 
holons and a bosonic operator $b_{i \sigma}$ describing neutral spin one-half 
bosonic spinons. Again, the no-double-occupancy constraint is imposed:
\be
\label{105}
 \sum_\sgm b_{i\sgm}^\dgr b_{i\sgm} + f_i^+f_i = 1\;  ,
\ee
so that the total number of holons and spinons is conserved,
$$
\sum_i \left ( \sum_\sgm \lgl b_{i\sgm}^\dgr b_{i\sgm} \rgl +
\lgl f_i^+ f_i \rgl \right ) = N \;  ,
$$   
being normalized to the total number of electrons. Here, the spinons are 
rather unphysical quasiparticles, being bosons with spin one-half. 

Fictitious quasiparticles, introduced by the slave particle representations, 
are not real particles, being just auxiliary tools for calculations. In these 
calculations, one has to preserve the imposed constraints that limit the
efficiency of the representations [143]. Under conditions, when the bosonic 
quasiparticles can be made conserved, they are allowed for formally experiencing 
condensation.

\section{Condensation of singletons and triplons}

Auxiliary quasiparticles are also introduced in the theory of quantum 
antiferromagnets, as suggested by Sachdev and Bhatt [144]. Generally, 
these quasiparticles can be defined either as bosons or as fermions. 
Treating them as bosons makes their condensation possible, provided 
they are conserved.

\subsection{Bond-operator representation}

Suppose there are two spins one-half, with the related operators
${\bf S}_1$ and ${\bf S}_2$. This pair of spins can be treated as a 
dimer connected by bonds. Sachdev and Bhatt [144] suggested for the pair 
of the spins the bond-operator representation
$$
S_1^\al = \frac{1}{2} \left ( s^\dgr t_\al + t_\al^\dgr s - 
i \sum_{\bt\gm} \ep_{\al\bt\gm} t_\bt^\dgr t_\gm \right ) \;   ,
$$
\be
\label{106}
S_2^\al = -\; \frac{1}{2} \left ( s^\dgr t_\al + t_\al^\dgr s + 
i \sum_{\bt\gm} \ep_{\al\bt\gm} t_\bt^\dgr t_\gm \right ) \;   ,
\ee
expressing the spins through the quasiparticle operator $s$, characterizing 
the singlet state, and the operators $t_\alpha$, with $\alpha = x,y,z$, 
describing the triplet state. Here, $\varepsilon_{\alpha \beta \gamma}$ is the 
totally antisymmetric unit tensor. The corresponding auxiliary quasiparticles
can, therefore, be called {\it singletons}, and {\it triplons}, respectively.
Explicitly, Eqs. (106) read as
$$
S_1^x = \frac{1}{2} \left [ s^\dgr t_x + t_x^\dgr s - 
i \left ( t_y^\dgr t_z - t_z^\dgr t_y \right ) \right ] \; ,
\qquad
 S_1^y = \frac{1}{2} \left [ s^\dgr t_y + t_y^\dgr s - 
i \left ( t_z^\dgr t_x - t_x^\dgr t_z \right ) \right ] \;   ,
$$
$$
S_1^z = \frac{1}{2} \left [ s^\dgr t_z + t_z^\dgr s - 
i \left ( t_x^\dgr t_y - t_y^\dgr t_x \right ) \right ] \; ,
$$
for the first spin, and as 
$$
S_2^x = -\;\frac{1}{2} \left [ s^\dgr t_x + t_x^\dgr s + 
i \left ( t_y^\dgr t_z - t_z^\dgr t_y \right ) \right ] \; ,
\qquad
 S_2^y = -\; \frac{1}{2} \left [ s^\dgr t_y + t_y^\dgr s +
i \left ( t_z^\dgr t_x - t_x^\dgr t_z \right ) \right ] \;   ,
$$
$$
S_2^z = -\;\frac{1}{2} \left [ s^\dgr t_z + t_z^\dgr s + 
i \left ( t_x^\dgr t_y - t_y^\dgr t_x \right ) \right ] \; ,
$$
for the second spin. This representation could be considered as a generalized 
Schwinger representation (96) extended to the pair of spins.

The restriction on the physical states to be either singlets or triplets 
leads to the constraint
\be
\label{107}
 s^\dgr s + \sum_\al t_\al^\dgr t_\al = 1 \;  ,
\ee
with $\alpha = x,y,z$.

The spin operators satisfy the standard commutation relations
\be
\label{108}
\left [ S_1^\al , \; S_1^\bt \right ] = 
i \sum_\gm \ep_{\al\bt\gm} S_1^\gm \; , \qquad 
\left [ S_2^\al , \; S_2^\bt \right ] = 
i \sum_\gm \ep_{\al\bt\gm} S_2^\gm \; , \qquad 
\left [ S_1^\al , \; S_2^\bt \right ] = 0
\ee
and are normalized as
\be
\label{109}
 \bS_1^2 = \bS_2^2 = S ( S + 1 ) = \frac{3}{4} \;  .
\ee

Under these conditions, the correct algebra of spin operators can be
reproduced by either bosonic or fermionic commutation relations for the 
operators of the auxiliary quasiparticles. The statistics of these 
quasiparticles is not uniquely defined; they can be either bosons or 
fermions. If one interprets them as bosons, then the bosonic 
commutation relations are assumed:
\be
\label{110}
 \left [ s , \; s^\dgr \right ] = 1 \; , \qquad 
\left [ t_\al , \; t_\bt^\dgr \right ] =  \dlt_{\al\bt} \; , 
\qquad 
\left [ s , \; t_\al \right ] = 
\left [ s , \; t_\al^\dgr \right ] = 0 \; .
\ee

The scalar product of two different spin operators is
\be
\label{111}
\bS_1 \cdot \bS_2 = \frac{1}{4} 
\left ( \sum_\al t_\al^\dgr t_\al - s^\dgr s \right )
\ee
and the summary $z$-component is
\be
\label{112}
 S_1^z + S_2^z = - i \left ( t_x^\dgr t_y - t_y^\dgr t_x \right ) \;  .
\ee

One introduces the following canonical transformation
\be
\label{113}
t_x = \frac{1}{\sqrt{2}} \left ( t_+ + t_- \right ) \; , 
\qquad
t_y = - \; \frac{i}{\sqrt{2}} \left ( t_+ - t_- \right ) \; , 
\qquad t_z = t_0 \;  .
\ee
The new operators satisfy the bosonic commutation relations
\be
\label{114}
\left [ t_\mu , \; t_\nu^\dgr \right ] = \dlt_{\mu\nu} \; , \qquad
[t_\mu , \; t_\nu ] = 0 \;   ,
\ee
where $\mu, \nu = +, -, 0$

The constraint (107) takes the form
\be
\label{115}
 s^\dgr s + \sum_\nu t_\nu ^\dgr t_\nu = 1 \; ,
\ee
with $\nu = +, -, 0$. The $z$-component sum (112) becomes
\be
\label{116}
 S_1^z + S_2^z = t_-^\dgr t_- - t_+^\dgr t_+ \;  .
\ee

\subsection{Lattice of dimers}

A lattice, consisting of two spin sublattices corresponding to two components,
can be treated as a lattice of dimers. Each pair of spins in a dimer can be
described as is explained in the previous subsection. Constraint (115) takes 
place at each lattice site enumerated by the index $j = 1,2,\ldots,N$,
\be
\label{117}
 s_j^\dgr s_j + \sum_\nu t_{j\nu}^\dgr t_{j\nu} = 1 \;  ,
\ee
where $\nu = +, -, 0$. This implies that the total number of quasiparticles
is conserved:
\be
\label{118}
 \sum_j \left ( \lgl s_j^\dgr s_j \rgl + 
\sum_\nu \lgl t_{j\nu}^\dgr t_{j\nu} \rgl \right ) = N \;  .
\ee

One usually assumes [144] that the singletons are completely condensed, so 
that $s_j$ can be replaced by the nonoperator real quantity $s$ equal to 
the average
\be
\label{119}
 s = \lgl s_j \rgl \;  ,
\ee
which does not depend on the site index because of the ideality of the 
lattice. 
  
Another standard assumption is that the main contribution among the triplons
is due to only one type of triplons denoted by 
\be
\label{120}
 t_j \equiv t_{j+} \;  .
\ee

Under these assumptions, the bond-operator representation reduces to
$$
S_{1j}^x = \frac{s}{2\sqrt{2}} \left ( t_j^\dgr + t_j \right ) =
- 2S_{2j}^x \; , \qquad 
S_{1j}^y = \frac{is}{2\sqrt{2}} \left ( t_j^\dgr - t_j \right ) =
- 2S_{2j}^y \; ,
$$
\be
\label{121}
 S_{1j}^z = - \; \frac{1}{2} \; t_j^\dgr t_j = S_{2j}^z \;  .
\ee
And the normalization condition (118) becomes
\be
\label{122}
 N s^2 + \sum_j \lgl t_j^\dgr t_j \rgl = N \;  .
\ee
Taking account of the lattice periodicity simplifies normalization (122) to
\be
\label{123}
s^2 + \lgl t_j^\dgr t_j \rgl = 1 \;   .
\ee
The number of singletons is defined self-consistently by requiring the 
system stability. Then triplons are conserved quasiparticles that can 
experience Bose-Einstein condensation.

\subsection{Order parameters}

There can exist two phases in the system, the normal phase of uncondensed 
triplons and that of condensed triplons [144]. The order parameter 
distinguishing these phases is the average $\langle t_j \rangle$. The normal 
phase, where triplons are not condensed, corresponds to magnetically disordered 
phase, while the condensed triplon phase is magnetically ordered [144]. The
second order parameter is the absolute value of magnetization 
\be
\label{124}
 M \equiv \frac{1}{N} \left | \sum_j \lgl S_{1j}^z + S_{2j}^z \rgl 
\right | \;  ,
\ee
reduced to the number of lattice sites. With representation (112), we have
$$
 M = \frac{1}{N} \left | \sum_j \lgl t_{jx}^\dgr t_{jy} - 
t_{jy}^\dgr t_{jx} \rgl \right | \;   .
$$
And representation (116) gives
$$
M = \frac{1}{N} \left | \sum_j \lgl t_{j+}^\dgr t_{j+} - 
t_{j-}^\dgr t_{j-} \rgl \right | \;   .
$$   
Resorting to simplification (121) yields
$$
S_{1j}^x = - S_{2j}^x \; , \qquad S_{1j}^y = - S_{2j}^y \; , \qquad
S_{1j}^z = S_{2j}^z \; , \qquad S_{1j}^z + S_{2j}^z = - t_j^\dgr t_j \; .
$$
Finally, using normalization (123), we get
\be
\label{125}
  M = \frac{1}{N} \sum_j \lgl t_j^\dgr t_j \rgl = 1 - s^2 \; .
\ee

\subsection{Disordered phase}

The system Hamiltonian was initially a functional of the spin operators
${\bf S}_{1j}$ and ${\bf S}_{2j}$. With the bond-operator representation 
(106), it became a functional of the quasiparticle operators $s_j$ and 
$t_{j \alpha}$, with $\alpha = x,y,z$. After the canonical transformation 
(113), the Hamiltonian became a functional of $s_j$ and $t_{j \nu}$, where 
$\nu = +,-,0$. With the following simplifications, resulting in Eqs. (121),
the Hamiltonian acquired the form of a functional of $s$ and $t_j$. The 
sequence of these transformations can be symbolically depicted as
\be
\label{126}
\hat H[\bS_{1j} , \; \bS_{2j} ] \longrightarrow 
\hat H [ s_j , \; t_{j\al} ] \longrightarrow 
\hat H[s_j , \; t_{j\nu} ] \longrightarrow 
\hat H [ s , \; t_j ] \;   .
\ee

In the phase, where triplons are not condensed, one has
\be
\label{127}
  \lgl t_j \rgl = 0 \; .
\ee
To guarantee the normalization condition (122), one has to define the grand
Hamiltonian
\be
\label{128}
 H = \hat H [ s , t_j ] - \mu \left ( Ns^2 + 
\sum_j t_j^\dgr t_j \right ) \;  ,
\ee
with the Lagrange multiplier $\mu$ being the system chemical potential. 
The number of singletons is defined by minimizing the system thermodynamic
potential, which leads to
\be
\label{129}
 \left \lgl \frac{\prt H}{\prt s} \right \rgl = 0 \;  .
\ee
This is a necessary condition for the system stability.

\subsection{Triplon condensation}

The Bose-Einstein condensation of triplons implies the appearance of
the order parameter
\be
\label{130}
 \eta = \lgl t_j \rgl \neq 0 \;  ,
\ee
which does not depend on the site index due to the lattice periodicity.
In order to correctly describe the condensed phase, we introduce the
Bogolubov shift
\be
\label{131}
 t_j = \eta + c_j \;  .
\ee
Thus, the Hamiltonian now is a functional of $s, \eta, c_j$, which means
the Hamiltonian transformation
\be
\label{132}
 \hat H [s, t_j ] \longrightarrow \hat H [ s, \; \eta , \; c_j ] \;  .
\ee  

The following numbers of quasiparticles are defined: the number of singletons 
\be
\label{133}
 N_s = \sum_j s^2 = Ns^2 \;  ,
\ee
the number of condensed triplons
\be
\label{134}
N_0 = \sum_j | \eta|^2 = N\eta^2 \; ,
\ee
and the number of uncondensed triplons
\be
\label{135}
 N_1 = \sum_j \lgl c_j^\dgr c_j \rgl = \lgl \hat N_1 \rgl \;  ,
\ee
where the number operator of uncondensed triplons is
$$
\hat N_1 \equiv \sum_j c_j^\dgr c_j \;   .
$$

The total number of quasiparticles, according to Eq. (123), is conserved:
\be
\label{136}
 N_s + N_0 + N_1 = N \;  .
\ee

To take into account these normalization conditions (133) to (135), it is
necessary to introduce the grand Hamiltonian with the required number of 
Lagrange multipliers, which defines a representative ensemble [16,18,145]. 
Following the general approach, as applied to Bose-condensed systems
[64-68], we need to introduce the grand Hamiltonian
\be
\label{137}
 H = \hat H [ s, \; \eta , \; c_j ] - \mu_s N_s -
\mu_0 N_0 - \mu_1 \hat N_1 \;  .
\ee
The Lagrange multipliers $\mu_s, \mu_0$, and $\mu_1$ guarantee the 
validity of the normalization conditions (133), (134), and (135),
playing the role of partial chemical potentials. The system chemical 
potential is defined by the variation of the system free energy $F$,
which gives
\be
\label{138}
 \mu = \frac{\prt F}{\prt N} = 
\frac{\prt F}{\prt N_s}  \frac{\prt N_s}{\prt N} + 
\frac{\prt F}{\prt N_0} \frac{\prt N_0}{\prt N} +
\frac{\prt F}{\prt N_1} \frac{\prt N_1}{\prt N} \; .
\ee
This, with the partial derivatives
\be
\label{139}
\mu_s = \frac{\prt F}{\prt N_s} \; , \qquad  
\mu_0 = \frac{\prt F}{\prt N_0} \; , \qquad
\mu_1 = \frac{\prt F}{\prt N_1} \; ,
\ee
yields the system chemical potential
\be
\label{140}
  \mu = \mu_s n_s + \mu_0 n_0 + \mu_1 n_1 \; ,
\ee
in which the quasiparticle fractions are defined as
\be
\label{141}
 n_s = \frac{\prt N_s}{\prt N} \; , \qquad 
n_0 = \frac{\prt N_0}{\prt N} \; , \qquad 
n_1 = \frac{\prt N_1}{\prt N} \; .
\ee
 
Not all partial Lagrange multipliers $\mu_s, \mu_0$, and $\mu_1$ are 
independent. Their relation follows from the condition of stability 
$\delta F = 0$, which  means
$$
 \frac{\prt F}{\prt N_s} \; \dlt N_s +  
\frac{\prt F}{\prt N_0} \; \dlt N_0 + 
\frac{\prt F}{\prt N_1} \; \dlt N_1 = 0 \; .
$$
In view of Eqs. (139), this is equivalent to
\be
\label{142}
 \mu_s \dlt N_s + \mu_0 \dlt N_0 + \mu_1 \dlt N_1 = 0 \;   .
\ee
Because of normalization (136), one has
\be
\label{143}
   \dlt N_s +  \dlt N_0 +  \dlt N_1 = 0 \;   .
\ee
Using this, Eq. (142) reduces to
$$
( \mu_0 - \mu_s ) \dlt N_0 + ( \mu_1 - \mu_s ) \dlt N_1 = 0
$$
and further to 
$$
 ( \mu_0 - \mu_s ) n_0 + ( \mu_1 - \mu_s ) n_1 = 0 \; .
$$
The latter gives
\be
\label{144}
 \mu_s = \frac{\mu_0 n_0 + \mu_1 n_1}{n_0 + n_1} \;  .
\ee
Substituting this $\mu_s$ into Eq. (140) results in
\be
\label{145}
 \mu = \mu_s \;  .
\ee
 
The quasiparticle fractions
\be
\label{146}
n_s = \frac{N_s}{N} = s^2 \; , \qquad n_0 = \frac{N_0}{N} \; , 
\qquad n_1 = \frac{1}{N} \sum_j \lgl c_j^\dgr c_j \rgl
\ee
satisfy the normalization
\be
\label{147}
 n_s + n_0 + n_1 = 1 \;  .
\ee
Hence, the fraction of condensed triplons is given by the equality
\be
\label{148}
 n_0 = 1 - n_1 - s^2 \;  .
\ee
Magnetization (125) is
\be
\label{149}
 M = 1 - s^2 = n_0 + n_1 \;  .
\ee

The stability conditions, requiring the minimization of the thermodynamic 
potential, lead to the equations
\be
\label{150}
\left  \lgl \frac{\prt H}{\prt n_s } \right \rgl = 0 \; , \qquad
 \left \lgl \frac{\prt H}{\prt n_0 } \right \rgl = 0 \; ,
\ee
where the multipliers $\mu_s, \mu_0, \mu_1$ are fixed. The first of these 
equations is equivalent to Eq. (129). Finally, the chemical potential 
$\mu_1$ is defined by the condition of the triplon condensate existence 
[16,18], which is equivalent to the requirement of a gapless spectrum, 
following from the Bogolubov [62] and Hugenholtz-Pines [69] theorems. 

In this way, all quantities are uniquely defined. It is worth stressing
the necessity of taking account of anomalous averages 
$\langle t_j t_j \rangle$. Neglecting these averages is mathematically 
unjustified, distorts thermodynamic relations, and spoils the phase 
transition order, making the Bose-Einstein condensation a first-order 
transition, as has been explained in Refs. [16,18,67,146]. While the 
proper account of the anomalous averages preserves the correct second 
order of the phase transition for Bose atoms [16,18,147,148], as well 
as for triplons [149,150].

Another principal point is the necessity of employing the representative 
ensemble with the grand Hamiltonian (137) in order to uniquely define a 
stable Bose-condensed system. For this, one has to introduce in the 
Hamiltonian the terms with the appropriate Lagrange multipliers. 
Forgetting some of these terms makes the system unstable. The majority 
of articles on triplon condensation do not use the correct grand 
Hamiltonian, hence, consider unstable systems.  

Let us emphasize that condition (129) fixes the number of singletons. As
far as the total number of all quasiparticles $N$ has been fixed from the 
beginning, being just the number of lattice sites, the number of triplons 
is  also fixed. Thence, triplons are conserved quasiparticles and can 
condense. Being conserved quasiparticles, triplons are principally 
different from unconserved magnons. The latter cannot experience Bose
condensation in an equilibrium system.

It is also important to remember that triplons are auxiliary quasiparticles
that, strictly speaking, are not precisely defined, since in the 
bond-operator representation (106), they can be considered either as bosons 
or as fermions. Choosing them as bosons is an arbitrary decision. They could 
equivalently be chosen as fermions. So, the introduction of bosonic triplons
is just a mathematical trick convenient for calculations, but the physical
nature of these triplons as bosons is rather ambiguous. Contrary to this, 
the definition of magnons through the Holstein-Primakoff representation (87) 
uniquely prescribes to magnons the Bose-Einstein statistics. Magnons are 
bosons by their birth. And magnons describe the straightforward physical 
reality characterizing spin oscillations. 

Unfortunately, many authors confuse triplons and magnons and write about 
magnon condensation in equilibrium dimer lattices. Of course, such a 
terminology, as has been stressed by Mills [36], is basically wrong. 
Triplons have nothing to do with magnons. The former are conserved 
quasiparticles and can condense in equilibrium, while the latter are not 
conserved and can never display equilibrium condensation.

\section{Strongly nonequilibrium condensates}

Unconserved quasiparticles cannot condense in equilibrium, although some
of them can condense in nonequilibrium systems, e.g., excitons, polaritons,
and photons. Coherent motion of transverse spins can also be interpreted as 
nonequilibrium magnon condensation. However, one should keep in mind that
the Holstein-Primakoff transformation is not valid for strongly 
nonequilibrium spin systems, because of which it cannot not be used for the 
latter. 

Nonequilibrium is favorable for the possibility of condensation of 
unconserved quasiparticle. But, contrary to this, nonequilibrium perturbations 
usually destroy the condensate of conserved particles. In this regard, it is 
illustrative to study what happens with the Bose system, prepared in the 
Bose-condensed state, after the system is subject to the action of 
increasing external perturbation driving the system far from its equilibrium 
state.

\subsection{Coherent topological modes}

Trapped atoms provide a very convenient object for studying well controlled  
transformations of an equilibrium Bose-condensed system into a highly excited 
state. There are different ways of exciting a Bose-condensed system by external 
fields. All of these ways can be mathematically classified into two categories.
One way is to act on the atomic density by applying an external field that is
characterized by a time-dependent potential $V({\bf r}, t)$. Another method is
to make the atomic scattering length $a_s(t)$ a function of time by invoking the 
Feshbach resonance techniques. Generally, the scattering length can be made a 
function of time and space. These two methods lead to similar consequences. 

When the amplitude of the perturbing field is small, elementary collective 
excitations are produced, representing density fluctuations around the given
equilibrium state. For short wavelengths $\lambda$, much shorter that the 
effective system size $L$, collective excitations in a nonuniform system are 
described by the Bogolubov spectrum
$$
 \ep(\bk,\br) = 
\sqrt{ c^2(\br) k^2 +\left ( \frac{k^2}{2m} \right )^2 } \;  ,
$$ 
where $c({\bf r})$ is local sound velocity. For long waves, such that 
$\lambda \gg L$, the spectrum of collective excitations is discrete and
depends on the trap geometry [1-12]. These elementary collective excitations
do not change the condensate fraction. 

But even rather weak perturbations may drive the system far from 
equilibrium, if the external field alternates with a frequency that is in 
resonance with a transition frequency between two energy levels of the trapped 
system. Recall that, for a trapped system, the spectrum of the stationary 
Gross-Pitaevskii equation is discrete and, because of atomic interactions,
it is not equidistant. This allows one to tune the frequency of the external
alternating field close to one of the transition frequencies.      

Coherent topological modes are the solutions of the stationary 
condensate-function equation [16,18,43,53,54]. At temperatures close to zero 
and asymptotically weak interactions, the condensate-function equation reduces 
to the stationary Gross-Pitaevskii equation
\be
\label{151}
 \hat H [ \vp_n ] \vp_n(\br) = E_n \vp_n(\br ) \;  ,
\ee
with the Gross-Pitaevskii Hamiltonian 
\be
\label{152}
 \hat H[\vp] = - \; \frac{\nabla^2}{2m} + U(\br) + 
N_0 \Phi_0 |\vp|^2 \;  ,
\ee 
in which $U({\bf r})$ is a trapping potential, $N_0$ is the number of condensed 
atoms, and 
$$
\Phi_0 \equiv 4\pi \; \frac{a_s}{m}
$$
is interaction strength, with $a_s$ being scattering length. The modes are 
called coherent, since they correspond to the Bose-condensed part of the system, 
which, by definition, is coherent. And they are termed topological because the 
corresponding atomic density is topologically different from the density of the 
ground state. 

Suppose that, in addition to the trapping potential, the system is subject to
the action of an alternating external potential
\be
\label{153}
 V(\br,t) = V_1(\br) \cos(\om t) + V_2(\br) \sin(\om t) \;  ,
\ee
whose frequency is tuned close to a chosen transition frequency
\be
\label{154}
  \om_{21} \equiv E_2 - E_1 \; .
\ee
This implies that the resonance condition 
\be
\label{155}
\left | \frac{\Dlt\om}{\om_{21}} \right | \ll 1 \qquad 
(\Dlt\om \equiv \om - \om_{21} )
\ee
is valid. The problem is analogous to that of resonance transitions of atoms,
with the principal difference in its nonlinearity caused by atomic interactions.  

Dynamics of the system is described by the time-dependent condensate-function 
equation [16,18,53] or, at zero temperature and weak interactions, by the 
Gross-Pitaevskii equation that is nothing but a nonlinear Schr\"{o}dinger 
equation
\be
\label{156}
 i\; \frac{\prt}{\prt t} \; \vp(\br,t) = \{ \hat H[\vp] + 
V(\br,t) \} \vp(\br,t) \;  .
\ee
The solution to this equation can be represented in the form of the expansion
over the coherent modes,
\be
\label{157} 
 \vp(\br,t) = \sum_n c_n(t) \vp_n(\br) e^{-iE_n t} \;  .
\ee
The fractional mode populations are given by the expressions
\be
\label{158}
p_n(t) \equiv | c_n(t) |^2 \;   .
\ee

Under the resonance condition (155), it is feasible to excite the coherent 
modes employing rather weak external fields. Similarly, by several alternating 
fields, one can generate several coherent modes [51,52]. If we increase the
amplitude of the perturbing external fields, then the topological coherent modes,
in addition to the direct resonance, with $\omega = \omega_{21}$, can also be 
generated by other processes, such as {\it harmonic generation} under condition
\be
\label{159}
n\om = \om_{21} \qquad ( n = 2,3,\ldots ) \;   ,
\ee
{\it parametric conversion}, when
\be
\label{160}
 \om_1 \pm \om_2 = \om_{21} \;  ,
\ee
and other {combined resonances}, when
\be
\label{161}
  \sum_i n_i \om_i = \om_{21} \qquad 
( n_i = \pm 1 , \pm 2 , \ldots) \; .
\ee
Thus, the increasing perturbation can generate a variety of different coherent 
modes. And, with a sufficiently strong perturbation, strict resonance conditions 
are not required.    

Among all these modes, there are quantized vortices. Their generation in 
a trap, perturbed by external fields, can be done so as to produce vortices 
with a chosen circulation, or the pairs of vortices and antivortices can be 
generated. One often employs cylindrical traps characterized by a transverse 
frequency $\omega_\perp$ and longitudinal frequency $\omega_z$, with the aspect 
ratio 
\be
\label{162}
 \al \equiv \frac{\om_z}{\om_\perp} = \left ( \frac{l_\perp}{l_z} 
\right )^2 \;  ,
\ee
where the related oscillator lengths are
$$
 l_\perp \equiv \frac{1}{\sqrt{m\om_\perp} } \; , \qquad
l_z \equiv \frac{1}{\sqrt{m\om_z} } \;  .
$$
The strength of atomic interactions can be described by the dimensionless
{\it coupling parameter}
\be
\label{163}
 g \equiv 4\pi N \; \frac{a_s}{l_\perp} \;  .
\ee
 
The basic vortex, having unit circulation [4], possesses the energy
\be
\label{164}
\om_{vor} = \frac{5}{2mR_{TF}^2} \; 
\ln \left ( 0.7 \; \frac{R_{TF}}{\xi} \right ) \;   ,
\ee
where the notation is used for the Thomas-Fermi radius squared
$$
R^2_{TF} \equiv \frac{1}{m\om_\perp} 
\left ( \frac{15}{4\pi} \; \al g \right )^{2/5}
$$
and the healing length
$$
 \xi \equiv \frac{1}{\sqrt{2m\rho\Phi_0} } \;  .
$$
The vortex energy can be written in the form
\be
\label{165}
 \om_{vor} = \frac{0.9\om_\perp}{(\al g)^{2/5} } \;
\ln ( 0.8 \al g) \;  .
\ee
 
The number of atoms $N$ in a trap is usually so large that the coupling 
parameter (163) is a large number $g \gg 1$. Then the dependence of the
coherent-mode transition frequencies on the coupling parameter is such 
that
$$
 \om_{mn} \; \propto \; ( \al g)^{2/5} \;  ,
$$
hence, these energies increase with $g$. While the energy of the basic 
vortex (165) decreases with $g$. This is why the basic vortex is the most 
energetically stable among all coherent modes, so that rather intensive
perturbation creates mainly such basic vortices [5,18]. The core of a 
vortex is the region with condensate depletion. Thence, vortex generation 
depletes the condensate fraction.

\subsection{Creation of separate vortices}

When the strength of external perturbations continues growing, this pumps
into the system additional energy. The action of an alternating potential
$V({\bf r}, t)$ increases the system energy so that the injected energy 
per atom is
\be
\label{166}
E_{inj} = \frac{1}{N} \int \rho(\br,t) 
\left | \frac{\prt V(\br,t)}{\prt t} \right | \; d\br dt \;  .
\ee 
Let the pumping potential have the amplitude $A$ and frequency $\omega$.
Then the injected energy (166), during the time interval $[t,t^\prime]$, 
approximately is
\be
\label{167}
 E_{inj} \approx A \om(t - t') \; .
\ee

A single vortex is created when the injected energy is of the order of 
the vortex energy, $E_{inj} \sim \omega_{vor}$, which implies
$$
A\om(t - t_0) \sim \om_{vor} \; .
$$
Therefore the crossover line of a single vortex creation, on the 
amplitude-time plane, is given by the expression
\be
\label{168}
 A_{vor} \; \sim \; \frac{\om_{vor} }{\om(t-t_0) } \;  .
\ee

Several vortices, of the number $N_{vor}$, can be generated when the 
injected energy reaches the value $E_{inj} \sim N_{vor} \omega_{vor}$, 
hence when
$$
 A\om(t - t_1) \; \sim \; N_{vor} \om_{vor} \;  .
$$
The more vortices are produced, the more condensate fraction is depleted,
though the system remains yet superfluid.

\subsection{Regime of quantum turbulence}

Producing more and more vortices (and antivortices), one can reach the 
state where the vortices form a random tangle. Such a random tangle of 
vortices signifies the appearance of the state of {\it quantum turbulence} 
[151-158]. The number of vortices, when the random tangle is being formed 
can be estimated [18] as 
\be
\label{169}
 N_{vor} \; \sim\; \frac{l_0}{\xi} \;  ,
\ee
where the effective trap parameters are
$$
 l_0 \equiv \frac{1}{\sqrt{m\om_0} } = 
\left ( l_\perp^2 l_z \right )^{1/3} \; , \qquad 
\om_0 = \frac{1}{m l^2_0 } = 
\left ( \om_\perp^2 \om_z \right )^{1/3} \; .
$$
This defines the crossover line of the starting turbulence
\be
\label{170}
 A_{tur} \; \sim \; \frac{l_0\om_{vor} }{\xi\om(t-t_1) } \;  .
\ee
In the turbulent regime the condensate fraction is getting more and more 
depleted. But the system can yet be treated as superfluid.

\subsection{Spatial condensate granulation}

With the developed turbulence, the condensate fraction becomes small. And 
finally, the condensate breaks into spatially separated pieces, or grains,
that are immersed into the normal phase without condensate. Then locally, 
in each grain, there is condensate. However, since the grains are spatially 
separated, superfluidity through the whole volume cannot arise. Generally, 
such a state is called {\it heterophase}, as far as it is formed by a mixture 
of different phases, randomly distributed in space [145,159-162]. In the 
present case, it can be termed {\it granulated condensate} [163]. The 
typical size of grains with condensate is given by the localization length  
\be
\label{171}
 l_{loc} = \frac{1}{m^2E_{inj}^2 l_0^3} = 
\left ( \frac{\om_0}{E_{inj}} \right )^2 l_0 \;  .
\ee
These grains are mesoscopic in size, in the sense that the above localization 
length is larger than the mean interatomic distance $a$, but shorter than the
effective trap size:
$$
 a < l_{loc} < l_0 \;  .
$$

The granulation starts when the injected energy is so large that the 
localization length becomes of the order of the trap size, 
\be
\label{172}
 l_{loc} \sim l_0 \; , \qquad E_{inj} \sim \om_0 \; .
\ee
This defines the crossover line of granulation
\be
\label{173}
A_{het} \sim \frac{\om_0}{\om(t-t_2)} \; .
\ee

Increasing the injected energy diminishes the grain size (171). And the 
condensate is completely destroyed, when the localization length (171) 
reduces to the order of the mean interatomic distance:
\be
\label{174}
 l_{loc} \sim a \; , \qquad 
E_{inj} \sim \om_0 \sqrt{\frac{l_0}{a} } \;  .
\ee
The corresponding crossover line
\be
\label{175}
 A_{nor} \sim \frac{\om_0}{\om(t-t_3)} \; \sqrt{\frac{l_0}{a} } 
\ee
shows where the granulated condensate is destroyed and the atomic cloud 
becomes a normal system without condensate. Since this system is strongly 
nonequilibrium, it can be named {\it chaotic fluid}.

\subsection{Amplitude-time phase diagram}

The above discussion can be summarized with the phase diagram in Fig. 1 
on the plane of the pumping field amplitude versus pumping time. The 
sequence of these states, except the chaotic fluid, has been observed in 
experiments [164-167]. The state of chaotic fluid has not yet been reached.
The details for the phase diagram can be found in Refs. [166,167].

In experiments [164-167], trapped atoms of $^87$Rb have been treated, 
having mass $m = 1.445 \times 10^{-22}$ g and scattering length 
$a_s = 0.577 \times 10^{-6}$ cm. The cloud was cooled down to low 
temperatures, much lower that the critical temperature of condensation
$T_c = 276$ nK, so that almost all $N = 2 \times 10^5$ atoms were condensed.
A cylindrical trap was used with the trap characteristics 
$$
\om_\perp = 2\pi \times 210\; {\rm Hz} = 
1.32 \times 10^3 {\rm s}^{-1} , \quad
\om_z = 2\pi \times 23\; {\rm Hz} = 
1.45 \times 10^2 {\rm s}^{-1} , \quad 
\om_0 = 0.63 \times 10^3 {\rm s}^{-1} ,
$$
$$
l_\perp = 0.74\times 10^{-4} {\rm cm} \; , \qquad 
l_z = 2.25\times 10^{-4} {\rm cm} \; , \qquad
l_0 = 1.08\times 10^{-4} {\rm cm} \; .
$$
This is an elongated trap with the aspect ratio 
$\alpha \equiv \omega_z/ \omega_\perp = 0.11$.

The effective atomic volume $V_{eff}$, average density $\rho$, mean 
interatomic distance $a$, and gas parameter $\gamma$, respectively, are
$$
V_{eff} \equiv \pi l_\perp^2 2 l_z = 2\pi l_0^3 = 
0.78\times 10^{-11} {\rm cm}^3 \; , \qquad
\rho \sim 2.55 \times 10^{15} {\rm cm}^{-3} \; ,
$$
$$
a \equiv \rho^{-1/3} = 0.73 \times 10^{-5} {\rm cm} \; , \qquad
\gm \equiv \rho^{1/3} a_s = \frac{a_s}{a} = 0.079 \;   .
$$
The dimensionless coupling parameters are
$$
 g = 4\pi N \; \frac{a_s}{l_\perp} = 1.96\times 10^4 \; , \qquad
\al g = 2.16 \times 10^3 \;  ,
$$
which implies strong effective atomic interactions. 

The trapped system of condensed atoms was subject to the action of a 
modulating field of frequency 
$$
\om = 2\pi \times 200\; {\rm Hz} = 1.26 \times 10^3 {\rm s}^{-1} \; ,
$$
hence, with the modulation period 
$T_{mod} \equiv 2 \pi / \omega = 5 \times 10^{-3}$ s. The action of this 
perturbation lasted during the modulation time $t_{mod}=0.02 - 0.06$ s.

The time of local equilibration is
$$
t_{loc} = \frac{m}{\hbar\rho a_s} = 
0.93 \times 10^{-4} {\rm s} \;   .
$$
Then the relation between the characteristic times is
$$
t_{loc} \ll T_{mod} \ll t_{mod} \;   .
$$
This means that the atomic system, during the process of modulation, has 
always been in the state of local equilibrium.

The healing length, defining the vortex core, is
$$
\xi \equiv \frac{1}{\sqrt{8\pi\rho a_s} } =
0.52 \times 10^{-5} {\rm cm} \; .
$$
Therefore the characteristic lengths are related as
$$
 a_s \ll \xi \sim a \ll l_0 \;  .
$$
The vortex energy is $\omega_{vor} = 0.41 \times 10^3$ s$^{-1}$. The 
turbulent regime starts when the number of the generated vortices reaches
the value
$$
 N_{vor} = \frac{l_0}{\xi} \approx 20 \; ,
$$ 
which is in good agreement with experiment [166,167].

The consideration of the present section illustrates how the system of 
conserved atoms, initially prepared in an almost pure Bose-condensed state, 
and being subject to the perturbation by external alternating fields, goes 
through a sequence of nonequilibrium regimes, ending with a normal chaotic 
fluid without condensate. The initial condensate fraction has been almost 
$100 \%$, while the final state contains no condensate. Nonequilibrium 
external perturbation depletes the condensate fraction of conserved particles.

A challenging problem would be to investigate experimentally the opposite 
process of equilibration of an excited atomic system, from the state of 
chaotic fluid back to the Bose-condensed state. Such problems of 
equilibration of finite quantum systems attract now high attention,
as can be inferred form the review articles [168-170].

\section{Conclusion}

For the phenomenon of the Bose-Einstein condensation, the distinction 
between particles and quasiparticles is secondary. The most important is 
whether their number is conserved or not. Conserved particles and 
quasiparticles can experience Bose-Einstein condensation in equilibrium
systems as well as can form nonequilibrium condensates. But unconserved 
quasiparticles cannot exhibit equilibrium condensation. However, they can
display condensation in nonequilibrium states, when an external pumping
creates a sufficient number of quasiparticles. Examples are nonequilibrium
condensates of excitons, polaritons, and photons.

It is possible to interpret the shape and Jahn-Teller phase transitions
as condensation of phonons. However this interpretation is useful solely
at the phase transition points, where the system is actually unstable. 
After the phase transition has occurred, there are no condensed phonons, 
but it is just necessary to redefine the mean atomic locations. It is
possible to avoid talking on phonon condensation by interpreting shape 
transitions as a sharp variation of the mean atomic locations. 
Self-consistently defined phonons never condense, whether in equilibrium
or nonequilibrium systems.

Among quasiparticles, there are those that enjoy direct physical meaning,
such as phonons and magnons. And there are auxiliary quasiparticles, 
appearing in the process of mathematical transformations and having no 
precise physical meaning, for instance, Schwinger bosons, slave bosons,
or singletons and triplons. These auxiliary quasiparticles even are not 
uniquely defined as bosons. If, by force, they are assumed to be bosons,
they can display equilibrium condensation, provided they are conserved.

When considering quasiparticle condensation, it is principally important 
to employ a correct terminology, not confusing conserved with unconserved 
quasiparticles. For example, triplons are auxiliary conserved quasiparticles
that are formally allowed to condense in equilibrium. And triplons must 
not be confused with magnons which are physical unconserved quasiparticles 
that cannot experience equilibrium condensation.   

The appearance of average transverse magnetization in strongly 
nonequilibrium spin systems can be interpreted as magnon condensation, 
though the Holstein-Primakoff transformation for such nonequilibrium
situations is not applicable.

To exhibit condensation, unconserved quasiparticles require nonequilibrium 
conditions. Contrary to this, the condensate of conserved particles becomes
depleted by nonequilibrium external perturbations.

\vskip 5mm

{\bf Acknowledgement}

\vskip 2mm
I am grateful for discussions to V.S. Bagnato, A. Garbaly, and E.P. Yukalova. 
Financial support from the Russian Foundation for Basic Research is appreciated.

\newpage

\begin{center}
{\Large{\bf Figure Caption} }
\end{center}

\vskip 3cm

{\bf Fig. 1}. Qualitative phase diagram showing the sequence of states 
on the plane of the pumping field amplitude versus pumping time.

\newpage

\begin{figure}[h]
\centerline{\includegraphics[width=15cm]{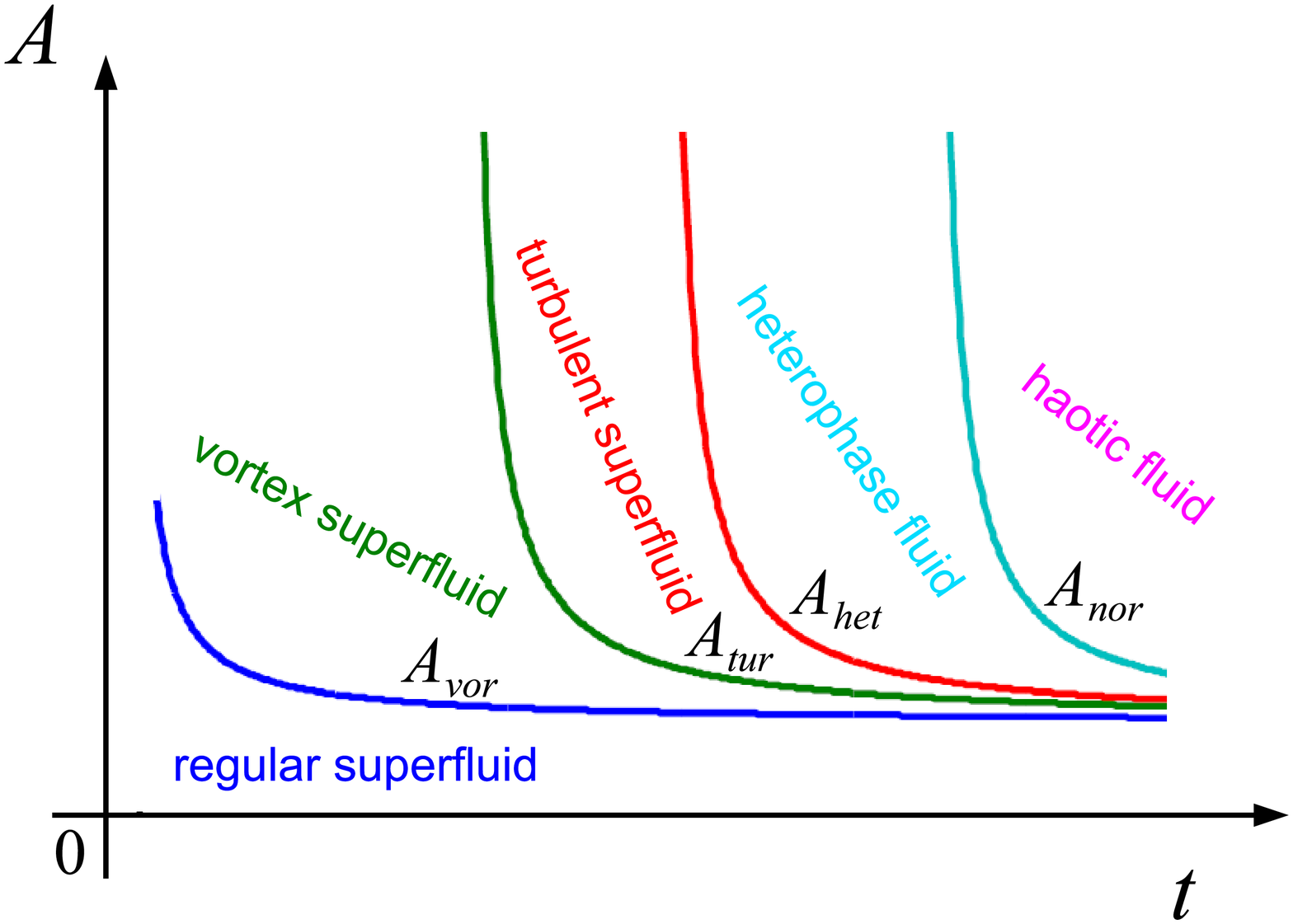}}
\caption{Qualitative phase diagram showing the sequence of states on the
plane of the pumping field amplitude versus pumping time.}
\label{fig:Fig.1}
\end{figure}

\end{document}